\documentclass[iop]{emulateapj}
\usepackage{bm}
\usepackage{multirow}
\usepackage{color}
\usepackage{float}
\makeatother
\newcommand\nar{\ref@jnl{NewAR}}
\makeatletter

\usepackage{natbib}
\usepackage{amsmath}
\usepackage{amsfonts}
\usepackage{amssymb}
\usepackage{epsfig}
\usepackage{graphicx}
\usepackage{booktabs}
\usepackage{array}
\usepackage{hyperref}
\hypersetup{colorlinks=true, citecolor=blue, filecolor=magenta,
 urlcolor=cyan, linkcolor=blue}

\shortauthors{Brown et al.}

\begin{document}	

\title{Infrared Contributions of X-Ray Selected Active Galactic Nuclei in Dusty Star-Forming Galaxies}

\author{Arianna Brown$^{1,2}$, Hooshang Nayyeri$^{1}$, Asantha Cooray$^{1}$, Jingzhe Ma$^{1}$, Ryan C. Hickox$^{3}$, and Mojegan Azadi$^{4}$ }
\affiliation{$^{1}$Department of Physics and Astronomy, University of California, Irvine, CA 92697}
\affiliation{$^{2}$Department of Physics and Astronomy, California State University, Los Angeles, CA 90032, USA} 
\affiliation{$^{3}$Department of Physics and Astronomy, Dartmouth College, Hanover, NH 03755, USA} 
\affiliation{$^{4}$Harvard-Smithsonian Center for Astrophysics, 60 Garden Street, Cambridge, MA, 02138, USA} 

\begin{abstract}
We investigate the infrared contribution from supermassive black hole activity versus host galaxy emission in the mid to far-infrared (IR) spectrum for a large sample of X-ray bright active galactic nuclei (AGN) residing in dusty, star-forming host galaxies. We select 703 AGN with $L_\text{X} = 10^{42}-10^{46}$\,ergs\,s$^{-1}$ at $0.1 < z < 5$ from the \textit{Chandra} XBo\"{o}tes X-ray Survey with rich multi-band observations in the optical to far-IR. This is the largest sample to date of X-ray AGN with mid and far-IR detections that uses spectral energy distribution (SED) decomposition to determine intrinsic AGN and host galaxy infrared luminosities. We determine weak or nonexistent relationships when averaging star-formation activity as a function of AGN activity, but see stronger positive trends when averaging $L_\text{X}$ in bins of star-forming activity for AGN at low redshifts. We estimate an average dust covering factor of 33\% based on infrared SEDs and bolometric AGN luminosity, corresponding to a Type 2 AGN population of roughly a third. We also see a population of AGN that challenge the inclination based unification model with individual dust covering factors that contradict the nuclear obscuration expected from observed X-ray hardness ratios. We see no strong connection between AGN fractions in the IR and corresponding total infrared, 24\,$\mu$m, or X-ray luminosities. The average rest-frame AGN contribution as a function of IR wavelength shows significant ($\sim80\%$) contributions in the mid-IR that trail off at $\lambda > 30\,\mu$m. Additionally, we provide a relation between observed $L_\text{X}$ and pure AGN IR output for high-z AGN allowing future studies to estimate AGN infrared contribution using only observed X-ray flux density estimates. 
\end{abstract}

\keywords{Galaxies: active --- Galaxies: nuclei --- Galaxies: evolution --- Infrared: galaxies --- X-rays: galaxies}

\section{Introduction}

Nearly all massive galaxies are believed to host a super-massive black hole (SMBH) at their center \citep{kormendy&richstone,magorrian, ho08}. Current research suggests that central black holes gain mass through a combination of both coalescence and bursts of mass accretion from the environment as the host galaxy evolves \citep[][and references therein]{kauf00,volonteri03,somerville,shankar, volonteri+}. The peak epoch of central black hole accretion, as the main source of active galactic nuclei (AGN), coincides with the peak epoch of star-formation in the universe at z$\approx 1-2$ \citep{dimat05, lutz08, aird10, stevens10, bonfield11, Alexander+Hickox}, and also major galaxy merger events \citep{dimat05, HH09, treister12,ellison2, rosario15}. Furthermore, in our local universe there exists a tight correlation between SMBH mass and host galaxy bulge mass and stellar velocity dispersions \citep[][and references therein]{ferra,marconi,gult,kormendy+}, whereas higher redshift SMBHs have been found in smaller host galaxies than expected \citep[e.g.][and references therein]{shields09}. These results signify that SMBH growth and galaxy growth are co-evolutionary processes and that these processes may regulate each other over time to produce the galaxy and SMBH sizes we observe today. 

Both central black hole growth and star formation rely on the abundance of cold molecular gas \citep{croton,smith08, dekel, dij09, bonfield11}. While cold dust and gas collapse to trigger star formation, the SMBH at the galaxy core gravitationally attracts cold gas and dust into a clumpy obscuring reservoir a few parsecs out from the SMBH, which fuels a thin, hot SMBH accretion disk with a radius typically $\lesssim 1$\,parsec \citep{antonucci,trist07,trist09,hopkins12,davies15}. The AGN feeds off the reservoir (hereby referred to as a torus; although it is now accepted that the dust is distributed in a more clumpy manner as opposed to a smooth donut structure \citep{nenkova1,nenkova2,sieb}) with a mass accretion process that emits X-ray, UV, and optical light \citep[e.g. see][]{haardt}. The X-ray, UV and optical light is partially absorbed by the surrounding dusty toroidal structure, then re-emitted in the infrared, making most AGN bright in the mid-IR, but not all AGN are X-ray bright \citep[e.g.][]{treister04,stern05,daddi2,donley12}. The current AGN unified model posits that AGN can be classified by the orientation of the dusty torus to the observer's line of sight \citep{antonucci,urry}: Type 1 AGN are usually observed face-on through a cavity in the torus and are typically bright in the X-ray, UV and optical spectrum; Type 2 AGN may be intrinsically less luminous or are observed at an angle through the torus, and are thereby obscured by high column densities of dust and gas ($N_{H} > 1.5 \times 10^{24}\,$cm$^{-2}$) from the observer's line-of-sight, enough so that most or all of the X-ray emission is absorbed and undetected \citep[e.g.][]{aird+2012,lanzuisi15}. However, recent observations are challenging this scheme \citep[e.g. see section 3.1 of][]{bian} and suggesting that observational differences in obscuration between AGN are mostly driven by individual SMBH accretion rates \citep[e.g.][]{lusso,ricci17} or host galaxy obscuration \citep[e.g.][]{goulding,netzer15,chen15,hickox2018}.

AGN accretion and outflow mechanisms are theorized to play a major role in galaxy evolution, via heating up, consuming and/or blasting away the host galaxy\textquoteright s remaining cold gas and dust necessary to create new stars, thereby triggering a star-formation quenching phase \citep[][and references therein]{dimat05,harrison14,hopkins06,fabian}. In observations, some AGN feedback processes are instantaneously strong enough to affect star formation in the host galaxy \citep[e.g.][but also see \citealt{leung17}]{sturm,reeves,rupke}; however, the exact contribution of the AGN phase to the physical properties of galaxies, compared to other mechanisms from stellar processes,  is still not well understood \citep[e.g.][]{silk,geach,diamond,gabor}, particularly for the most powerful AGN \citep[e.g.][]{stanley,Rosario+2012}. To study the effect of powerful AGN on their host galaxies, it is necessary to have a large statistical sample of AGN with multi-band observations to individually derive and constrain their physical properties.

One of the main degeneracies in determining the evolutionary relationship between AGN and host galaxy star-formation lies in their mutual obscuration by warm dust \citep{delve15,symeo16,lutz08}. The radiation originating from warm dust in stellar nebulae and from the obscuring torus around AGN are both bright in the mid to far-IR spectrum and thus necessary to disentangle prior to using IR radiation as an indicator for any host galaxy dust properties, including measurements of dust temperatures, host galaxy stellar mass, and star-formation rates; without this decomposition, there is a risk of measurement overestimation and, therefore, an increase in uncertainties. AGN accretion and outflow mechanisms release a large amount of energy detectable at nearly all wavelengths, in particular X-rays from the accretion disk \citep[see][for a review of AGN viewed in the X-ray spectrum]{brandt} and radio signatures from synchroton radiation \citep[e.g.][]{miley,blanford,jorstad,condon}. These features are the most commonly utilized as identifiable signatures that could be used to distinguish AGN from their host galaxies \citep{donley05,delmoro,mush,brandt}.

Observational studies and models of IR SEDs for local AGN reveal radiative flux densities that generally increase through the mid-IR then rapidly decline starting somewhere between $40\, \mu m < \lambda < 100\, \mu$m out to sub-millimeter wavelengths \citep{mullaney}. Prior to the \textit{Herschel Space Observatory} \citep{pilbrat}, observations were limited out to $ \lambda < 200\,\mu$m only for a small sample of very far-IR bright, mostly local objects \citep[e.g.][]{omont,haas}. \textit{Herschel} has been instrumental in constraining the dust SEDs for large samples of local and high redshift AGN and star forming galaxies, revealing a universe that is optically obscured by dust and therefore undetected at shorter wavelengths \citep[e.g.][and references therein]{mullaney15,symeo16,casey2014dusty}. 

\begin{figure}[t]
\centering\includegraphics[width=0.47\textwidth, height=7cm]{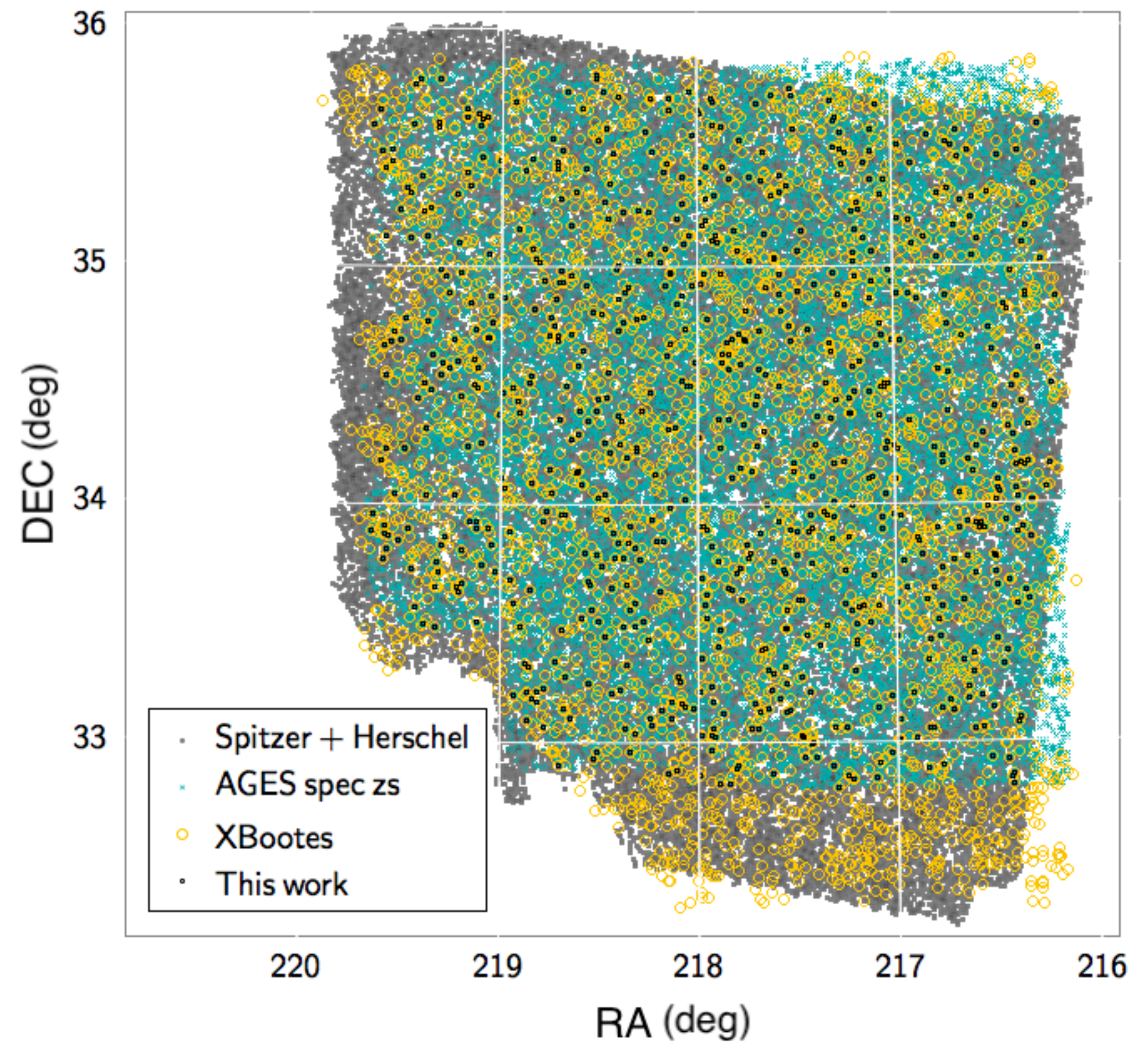}
\caption{Survey map for the parent surveys from which our main sample is derived. Grey points mark all galaxies with a S/N $>3$ in the MIPS 24$\mu m$ band and at least one SPIRE band; blue points denote all galaxies in the AGES survey with spectroscopic redshifts; gold circles outline all of the X-ray sources in the XBo\"{o}tes survey; black points mark our final sample of 703 AGN and host galaxies which spans $\sim 7$\,deg$^{2}$. Respective survey coverage and depths are discussion in Section \ref{data}.}
\label{coverage}
\end{figure}

In this paper, we use multi-wavelength infrared observations from the \textit{Herschel} Space Observatory \citep{spire+,pacs+} combined with the \textit{Spitzer} Space Telescope \citep{spitzer}, along with optical wide-area observations, and X-ray data from the \textit{Chandra} X-ray Observatory \citep{chandra} to construct the  AGN and host galaxy SEDs and explore the warm dust properties in the context of AGN accretion activity. We focus on X-ray selected AGN in the wide 9.3 deg$^{2}$ Bo$\ddot{o}$tes legacy field \citep{ndwfs} with mid and far-IR counterparts detected by \textit{Herschel} and \textit{Spitzer} \citep{oliver2012herschel,SDWFS}. The rich amount of data in the IR allows us to avoid the uncertainties that arise from single-band SED fitting. Furthermore, the multi-wavelength detections allow us to reliably use SED decomposition models to isolate AGN contribution in the infrared, reducing the likelihood of AGN contamination when estimating host galaxy properties. 

\begin{figure}[t]
\includegraphics[width=0.48\textwidth]{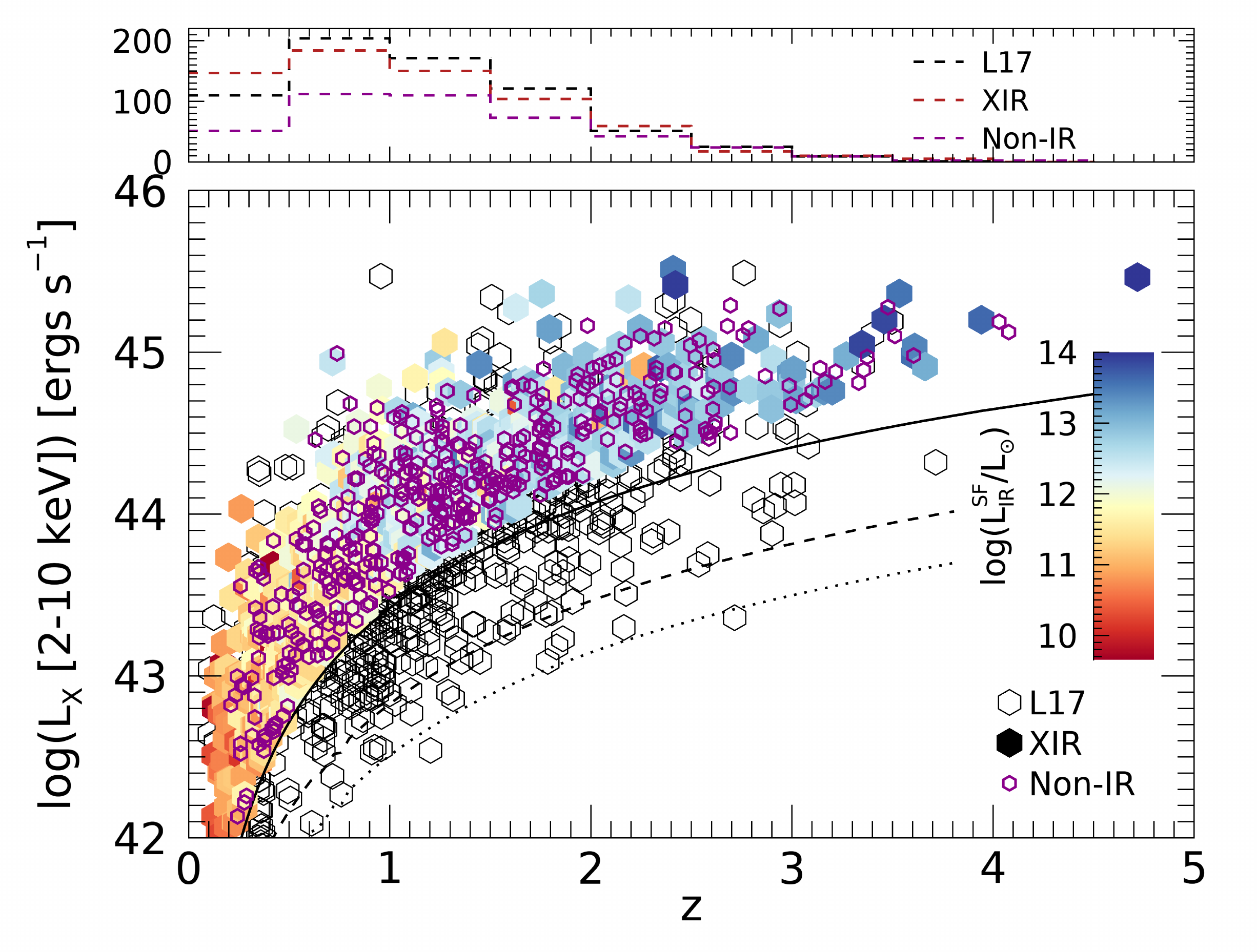}
\caption{Distribution of rest-frame X-ray luminosities and spectroscopic redshifts for our AGN sample. The solid circles are the 703 X-ray AGN with \textit{Spitzer} 24\,$\mu$m and far-infrared \textit{Herschel} detections. Colors represent rest-frame, infrared luminosities corrected for AGN contamination derived from individual respective SEDs (see Section \ref{samplesel}). The purple circles are the 425 X-ray AGN without mid/far-IR detections. The black empty circles are the AGN used for analysis in L17. The black solid line represents the X-ray flux limit of the \textit{Chandra} XBo\"{o}tes survey \citep{murray+2005}; for comparison, the dashed and dotted lines mark the sensitivity limits of the \textit{XMM-Newton} \citep{xmm} and \textit{Chandra} \citep{ccosmos} surveys in the COSMOS field, respectively. Shown in the top panel is the number of sources in our XIR sample (red), non-IR sample (purple), and in L17 (black) in redshift bins of size 0.5.}
\label{zvslx}
\end{figure}

This paper is organized as follows. Section \ref{data} describes the multi-wavelength survey data used in this analysis. Section \ref{samplesel} details the AGN sample selection procedure. In Section \ref{results}, we discuss the derivation of AGN and host galaxy properties and the results in the context of other published studies; section \ref{summary} provides a summary of this work. Throughout this study, we assume a cosmology with $H_{0}\,=70$\,km\,s$^{-1}$, $\Omega_{m}\,=\,0.3$, and $\Omega_{\Lambda}\,=\,0.7$.

\begin{figure}[h]
\centering\includegraphics[width=0.48\textwidth, height=10cm]{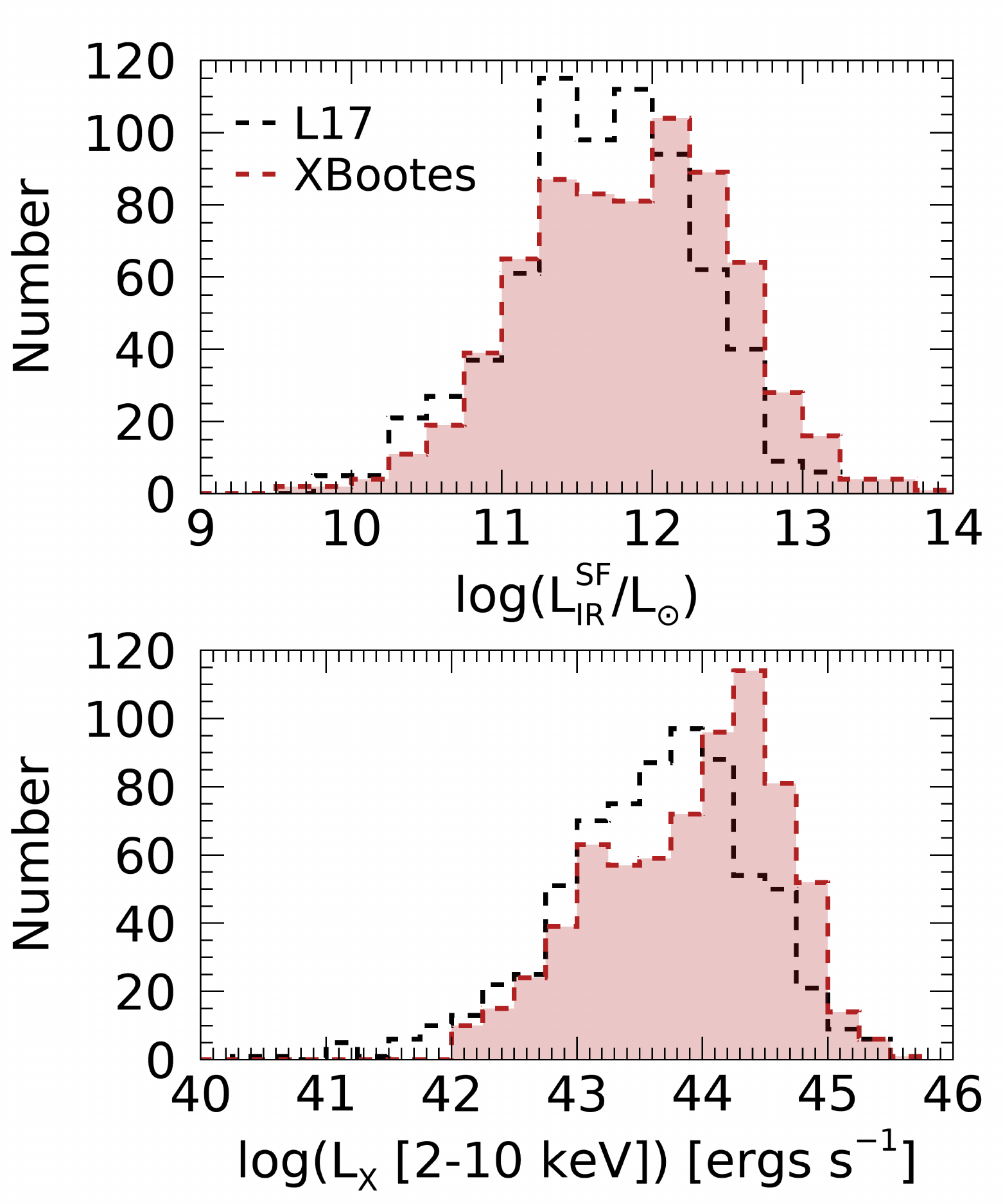}
\caption{Distributions of AGN and host-galaxy properties comparing this sample (red) and \citet{Lanzuisi} (L17; black) samples. \textit{Top}: Histogram of rest-frame, AGN-corrected infrared luminosities in bins of 1 dex, with median infrared luminosities of $1.95\times10^{45}$ ergs\,s$^{-1}$ and $2.69\times10^{45}$ ergs\,s$^{-1}$ for L17 and our sample, respectively. \textit{Bottom}: Histogram of rest-frame X-ray luminosities in bins of 1 dex; our sample has a slightly higher median X-ray luminosity of $L\mathrm{_{X}}=1.07\times10^{44}$\,erg\,s$^{-1}$ compared to the L17 median X-ray luminosity of $L\mathrm{_{X}}=4.79\times10^{43}$\,erg\,s$^{-1}$.}
\label{hist}
\end{figure}

\section{Multi-wavelength Data} \label{data}

The survey observations used in this study are centered in the Bo$\ddot{o}$tes field at $\alpha =$ 14h 30m 05.71s, $\delta = +$34$^{\circ}$ 16$^{\prime}$ 47$^{\prime\prime}$.5 \citep{ndwfs}. We use publicly available photometric catalogs ranging from optical to far-infrared wavelengths, complemented with X-ray data and spectroscopic redshifts, with known active galaxies \citep{SDWFS} and clusters of AGN \citep{brand2006chandra}. The multi-wavelength observations cover different areas across the Bo$\ddot{o}$tes field (see Figure \ref{coverage}). Table \ref{fluxtable} summarizes the data used and respective approximate field coverage.

The wide-area XBo$\ddot{o}$tes survey provides us with a unique opportunity to probe a large population of the most powerful AGN, half of which are also embedded in galaxy powerhouses with total infrared luminosities (L$\mathrm{_{IR}}$) greater than $10^{12}L_{\odot}$ (also known as ultra-luminous infrared galaxies or ULIRGs). Some weakly accreting AGN and AGN obscured by Compton thick hydrogen column densities ($N_{H} > 1.5 \times 10^{24}\,$cm$^{-2}$) may be missed by X-ray surveys \citep[e.g.][]{aird+2012,lanzuisi15}. However, studies confirm no single waveband can be used to select a complete sample of AGN, and X-ray detections remain one of the most reliable identification methods \citep[e.g.][and references therein]{barmby,mendez,ellison,cowley,azadi17,brandt}.

\renewcommand{\arraystretch}{1.25}

\begin{table*}[t]
\caption{Population counts and field coverage of the multi-wavelength flux catalogs used to generate 703 individual SEDs. 
}
\begin{center}
\begin{tabular}{m{16em} m{10em} m{10em} m{10em}}
\hline \hline 
Name & Bands & Survey Size & N Detected in Sample\\
\hline 
XBo\"{o}tes \citet{murray+2005} & 0.5-7\,keV & $\sim9.3$ deg$^{2}$ & 703 \\
NDWFS \citet{ndwfs} & $B_{w}$, $R$, $I$, and $K$ & $\sim9.3$ deg$^{2}$ & 652\\
IR Bo\"{o}tes Imaging Survey \citet{newfirm} & $H$ and $J$ & $\sim9.3$ deg$^{2}$ & $\sim$325\\
SDWFS \citet{SDWFS} & 3.6, 4.5, 5.8 and 8.0\,$\mu$m & $\sim10$ deg$^{2}$ & $\sim$330\\
HerMES MIPS \citet{oliver2012herschel} & 24\,$\mu$m & $\sim10$ deg$^{2}$ & 703\\ 
HerMES PACS \citet{oliver2012herschel} & 110 and 170\,$\mu$m & $\sim3$ deg$^{2}$ & 138 and 181\\
HerMES SPIRE \citet{oliver2012herschel} & 250, 300 and 500\,$\mu$m & $\sim8.5$ deg$^{2}$ shallow, $\sim3$ deg$^{2}$ deep & 489, 398, and 159\\ 
\hline
\end{tabular} 
\end{center}
\label{fluxtable}
\end{table*}

\subsection{X-ray Data}
Our AGN sample is selected from the \textit{Chandra} XBo\"{o}tes Survey, a 5-ks X-ray survey of the 9.3 deg$^2$ Bo\"{o}tes Field as defined in the NOAO Deep Wide-Field Survey \citep[NDWFS;][]{murray+2005}. This survey covers the full area defined by NDWFS with 126 individual 5\,ks contiguous pointings at uniform observational depths of $f_{0.5-7 \, \text{keV}} \sim $ $8 \times {10^{-15}}\,$erg\,s$^{-1}$\,cm$^{-2} $, yielding 3293 point sources with four or more counts. Rest-frame X-ray luminosities are determined by the following equation \citep{alexander03}: 

\begin{equation}
\mathrm{L_{X} = 4\pi \times D_{L}^{2} \times F \times (1 + z)^{\Gamma - 2}}
\end{equation}
 
\noindent where $D_{L}$ is the luminosity distance, $F$ is the hard band X-ray flux, $z$ is the redshift and a photon index of $\Gamma$ = 1.9, which is typical for an unabsorbed X-ray luminous AGN \citep[e.g.][]{vignali,nandra+}. To remain consistent in comparison to other studies, we translate our full band $0.5-7$ keV luminosities to $2-10$ keV hard band luminosities with a conversion factor of 0.78, which is the ratio of respective intensities over each keV energy range for $\Gamma$ = 1.9. Due to the shallow nature of the XBo\"{o}tes Survey, spectral fitting to correct for X-ray absorption is difficult or unachievable at an individual level for $\sim$90\% of our sources \citep[see][for a more detailed discussion]{kenter,murray+2005}, so we leave the observed fluxes to be interpreted at face value. We select sources with X-ray luminosities L$\mathrm{_{X}} > 10^{42}\,$erg\,s$^{-1}$ as lower luminosity sources may contain contamination from host galaxy processes \citep[e.g. supernovae, X-ray binaries and massive stellar outflows;][]{ranalli,mineo1,mineo2,lehmer}. 
  The X-ray survey depth of this study allows us to probe a larger population of the brighter end of the AGN luminosity function (see Figure \ref{zvslx}). Figure \ref{hist} (bottom) displays the X-ray population distribution of this sample. The wider coverage of the XBo\"{o}tes Survey allows us to study a large sample of powerful AGN with $50\%$ of the 703 selected sources residing at or above L$\mathrm{_{X}}$ $= 1.07 \times 10^{44}$\,ergs\,s$^{-1}$; similar studies using surveys that may be deeper but cover smaller areas in the sky yield populations of weaker AGN; for example, \citet{Lanzuisi} (L17, hereafter) analyzed 692 X-ray selected AGN in the COSMOS field \citep{cosmos} with a median L$\mathrm{_{X}} = 4.79 \times 10^{43}$\,ergs\,s$^{-1}$.

\subsection{Infrared Data}

Mid-IR and far-IR fluxes are collected from Data Release 4 of the \textit{Herschel} Multi-tiered Extragalactic Survey\footnote{http://hedam.oamp.fr/} \citep[HerMES;][]{oliver2012herschel}. Far-IR observations were taken by the \textit{Herschel} Spectral and Photometric Imaging Receiver (SPIRE) at 250\,$\mu$m, 350\,$\mu$m, and 500\,$\mu$m \citep{spire+}, and the \textit{Herschel} Photoconductor Array Camera and Spectrometer (PACS) $110$\,$\mu$m and $170$\,$\mu$m \citep{pacs+} bands; mid-IR observations were completed by the \textit{Spitzer} multi-band Imaging Photometer (MIPS) at 24\,$\mu$m \citep{mips+}. Fluxes for all five \textit{Herschel} bands used in the HerMES survey are recorded on positions defined by MIPS 24\,$\mu$m priors with a respective $5\sigma$ detection limit at $\sim 0.3$ mJy. The HerMES SPIRE campaign consisted of a combination of both deep and shallow observations: the center $\sim3$ deg$^{2}$ region is deeper and reaches $5\sigma$ detection limits at 13.8, 11.3, and 16.4\,mJy at 250\,$\mu$m, 350\,$\mu$m, and 500\,$\mu$m, respectively; the outer $\sim8.5$ deg$^{2}$ region surrounding the center reaches $5\sigma$ detection limits at 25.8, 21.2 and 30.8 mJy for the 250\,$\mu$m, 350\,$\mu$m and 500\,$\mu$m bands, respectively. The PACS observations occurred over the center $\sim3$ deg$^{2}$ of the Bo$\ddot{o}$tes region reaching 5$\sigma$ depths of 49.9 and 95.1\,mJy for the 110 and 170\,$\mu$m bands, respectively. Uncertainties in this analysis include both instrumental and confusion noise; we refer the reader to \citet{roseboom} for a more detailed description of flux uncertainty determinations in the HerMES catalogs.


Near/Mid-IR catalogs were compiled from the \textit{Spitzer} Deep, Wide-field Survey (SDWFS) \citep{SDWFS} which used all four channels of the \textit{Spitzer} Infrared Array Camera (IRAC) \citep{irac+} to image the entire $\sim10$ deg$^{2}$ Bo\"{o}tes field. SDWFS is a combined four epoch survey that contains $\sim10^{5}$ sources per band detected at 5$\sigma$ depths of 19.77, 18.83, 16.50, and 15.82 Vega mag at 3.6\,$\mu$m, 4.5\,$\mu$m, 5.8\,$\mu$m, and 8.0\,$\mu$m, respectively. We also use $J$ and $H$-band data from the NEWFIRM Infrared Bo\"{o}tes Imaging Survey \citep{newfirm} which reaches 5$\sigma$ limits of 22.05 and 21.30 Vega mag, respectively; and optical $B_{w}$, $R$, $I$ and $K$-band data from the NDWFS survey \citep{ndwfs} reaching 5$\sigma$ depths\footnote{https://www.noao.edu/noao/noaodeep/} at 26.6, 26.0, 26.0, and 21.4 AB mag, respectively. For all IR bands, we consider source detections at $ > 3\sigma$.

\begin{figure*}[!t]
\begin{tabular}{cc}
\includegraphics[width=0.5\textwidth]{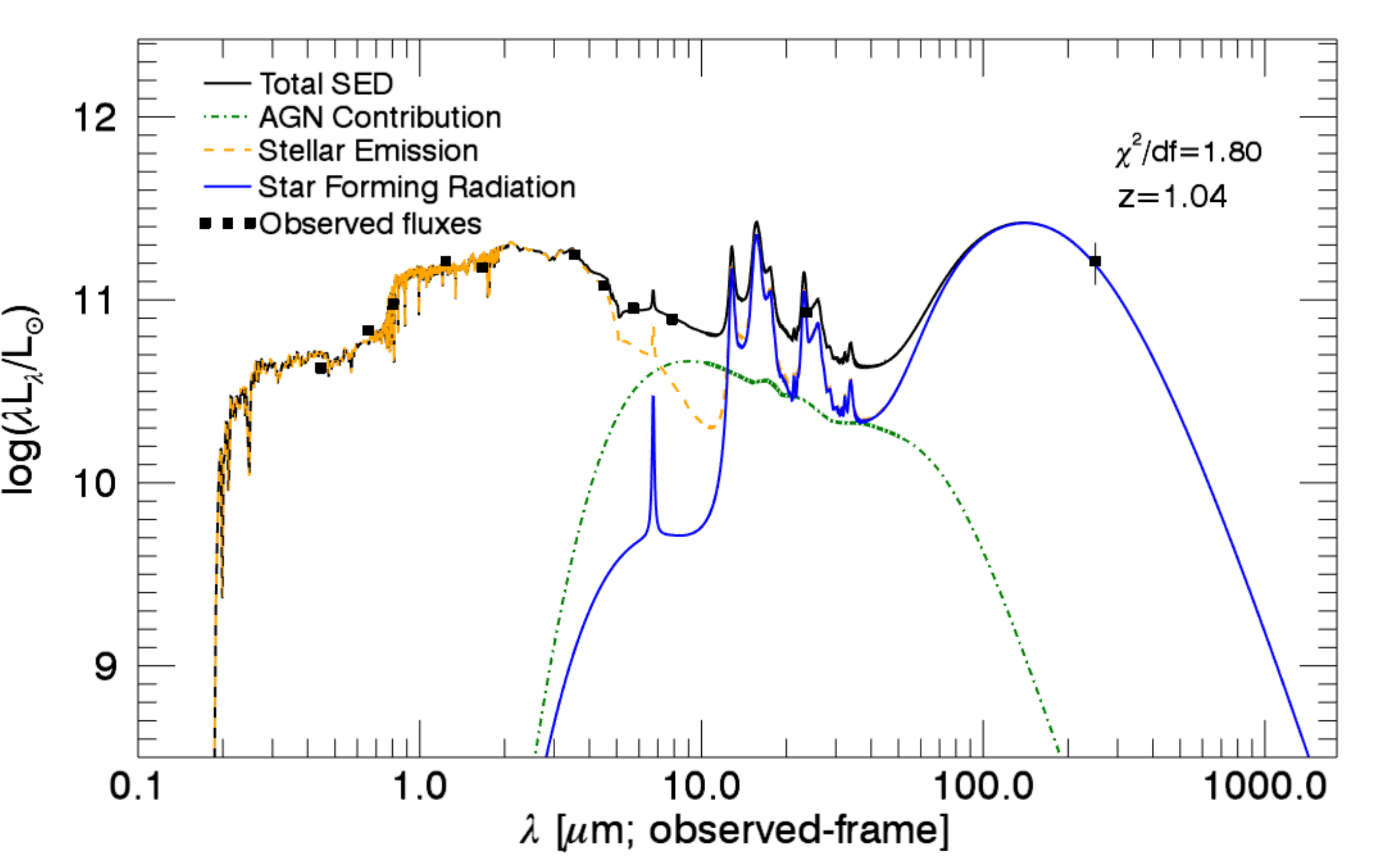}
&
\includegraphics[width=0.5\textwidth]{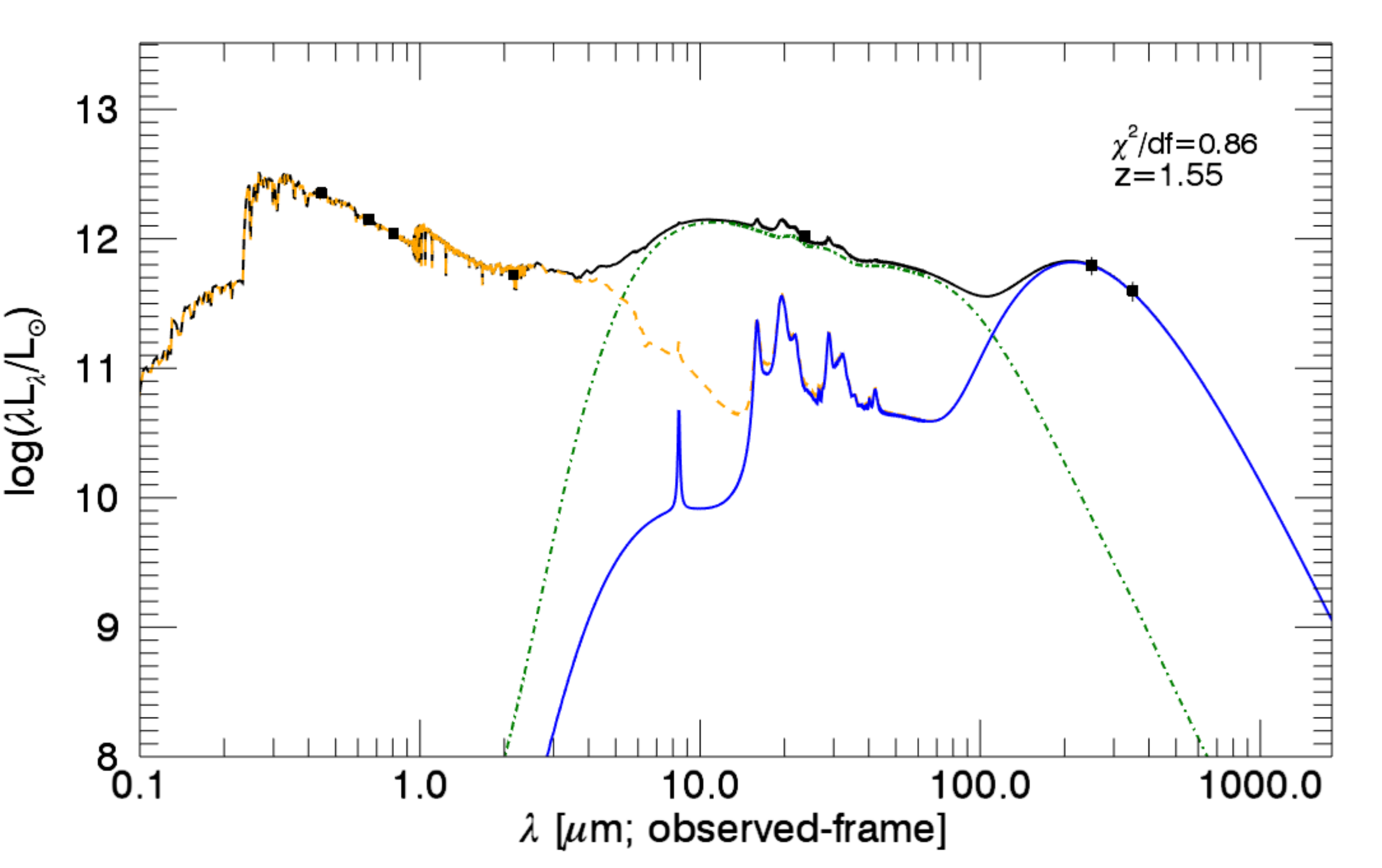}
\end{tabular}
\caption{Example spectral energy distributions generated by \textsc{sed3fit} \citep{berta}. The dashed gold line is the stellar emission contribution, the blue line is the radiation contributed by star formation processes, the green dashed line is the contaminating radiation from the AGN including the heated dusty torus surrounding the black hole, and the black line is the total SED or the summation of the three components.\textit{Left}: SED for a galaxy with star formation processes dominating the mid to far-infrared spectrum. \textit{Right}: In this SED, the AGN component provides the most contribution in the mid-infrared (and some of the far-IR) spectrum that would typically be attributed to star formation processes.}
\label{seds}
\end{figure*}

\subsection{Redshifts}
Spectroscopic redshifts are extracted from the AGN and Galaxy Evolution Survey (AGES) \citep{ages}, an optical spectroscopic and photometric redshift survey for optically selected sources in 7.7 deg$^{2}$ of the Bo\"{o}tes field. We limited our sample to spectroscopic redshifts in the range $z > 0.1$ (Figure \ref{zvslx}) to avoid the uncertainties associated with photometric redshifts and avoid contamination by local AGN and ULIRGs. 

To investigate the evolution of AGN and galaxy properties with redshift, we complete our analysis over five redshift intervals and consider the X-ray - infrared relationship in each respective interval. The following redshift intervals are designed so that each interval has a sufficient number of sources ($\sim90-200$) to create several statistically significant bins within that range: $z=0.1-0.4$, $0.4-0.8$, $0.8-1.2$, $1.2-2$, and $2-5$. These redshift bins (z-bins) are consistent in comparison with several other similar studies, and contain 95, 178, 140, 195, and 95 sources, respectively.

\section{AGN Sample Selection} \label{samplesel}

	The final sample used in this study consists of powerful AGN with spectroscopically confirmed redshifts, and a detection in one \textit{Herschel} SPIRE or PACS band. Since all objects in the HerMES campaign are based on \textit{Spitzer} MIPS priors, it follows that every object in our sample has at least one 24\,$\mu$m detection as well as one \textit{Herschel} detection. We achieve this sample, dubbed the XIR sample, through the following methods.

We matched X-ray AGN to infrared counterparts and spectroscopic redshifts using a nearest neighbor matching technique. First, X-ray sources were matched to the AGES redshift catalog using a 1$^{\prime\prime}$ search radius on their optical coordinates from \citet{brand2006chandra}, with a spurious match rate estimated at \textless 1\%. We were able to use such a small search radius confidently due to prior work by \citet{brand2006chandra} who used a Bayesian matching scheme to determine optical counterparts for 98\% of the X-ray sources in the XBo\"{o}tes survey under a 1$^{\prime\prime}$ search radius.  We note that AGES redshifts were determined using optical spectroscopy, and as such this study explores the properties of brighter, less dust obscured active galaxies. We also note that the AGES survey misses $\sim2$\,deg$^{2}$ of the XBo$\ddot{o}$tes and HerMES survey (Figure \ref{coverage}), removing $10\%$ of X-ray sources as possible candidates for this study. Near-IR and optical data were matched to the MIPS 24\,$\mu$m coordinates from the HerMES catalog \citep{oliver2012herschel} using a 3$^{\prime\prime}$ search radius, which corresponds to the \textit{Spitzer} MIPS 24\,$\mu$m half width at half maximum. Finally, we matched the MIPS 24\,$\mu$m coordinates to the AGES coordinates. Again, we estimate a spurious match rate of \textless 1\% when matching infrared data together, and once more when matching infrared data to X-ray sources with spectroscopic redshifts. 

Prior to fitting a spectral energy distribution, we require an object to have a 24\,$\mu$m detection and a detection in one of the \textit{Herschel} bands. The far-IR survey was defined on the coordinates for sources detected at 24\,$\mu$m, thus any Bo\"{o}tes source detected by \textit{Herschel} will also have a measurement at 24\,$\mu$m. Even though \textit{Herschel} observational depths varied across the inner and outer region of the survey area, we still find a uniform density of $\sim100$ AGN per square degree that satisfy our selection criteria. Additionally, due to the work by \citet{brand2006chandra}, the majority ($\sim 93\%$) of the sample also has an optical detection. 

The mid and far-IR photometry requirement is unique to this work. Comparable studies required only one mid or far-IR detection or relied on stacking techniques and photometric upper limits to supplement, creating large uncertainties when generating AGN SEDs, particularly on the Wien side of the far-IR SED corresponding to dust emission \citep[e.g.][]{mullaney12,stanley, Lanzuisi}. With the mid and far-IR requirement, we can better constrain dusty torus emission for powerful AGN and host star-forming galaxies. 

\begin{figure*}[!t]
\begin{tabular}{lll}
\includegraphics[width=0.32\textwidth]{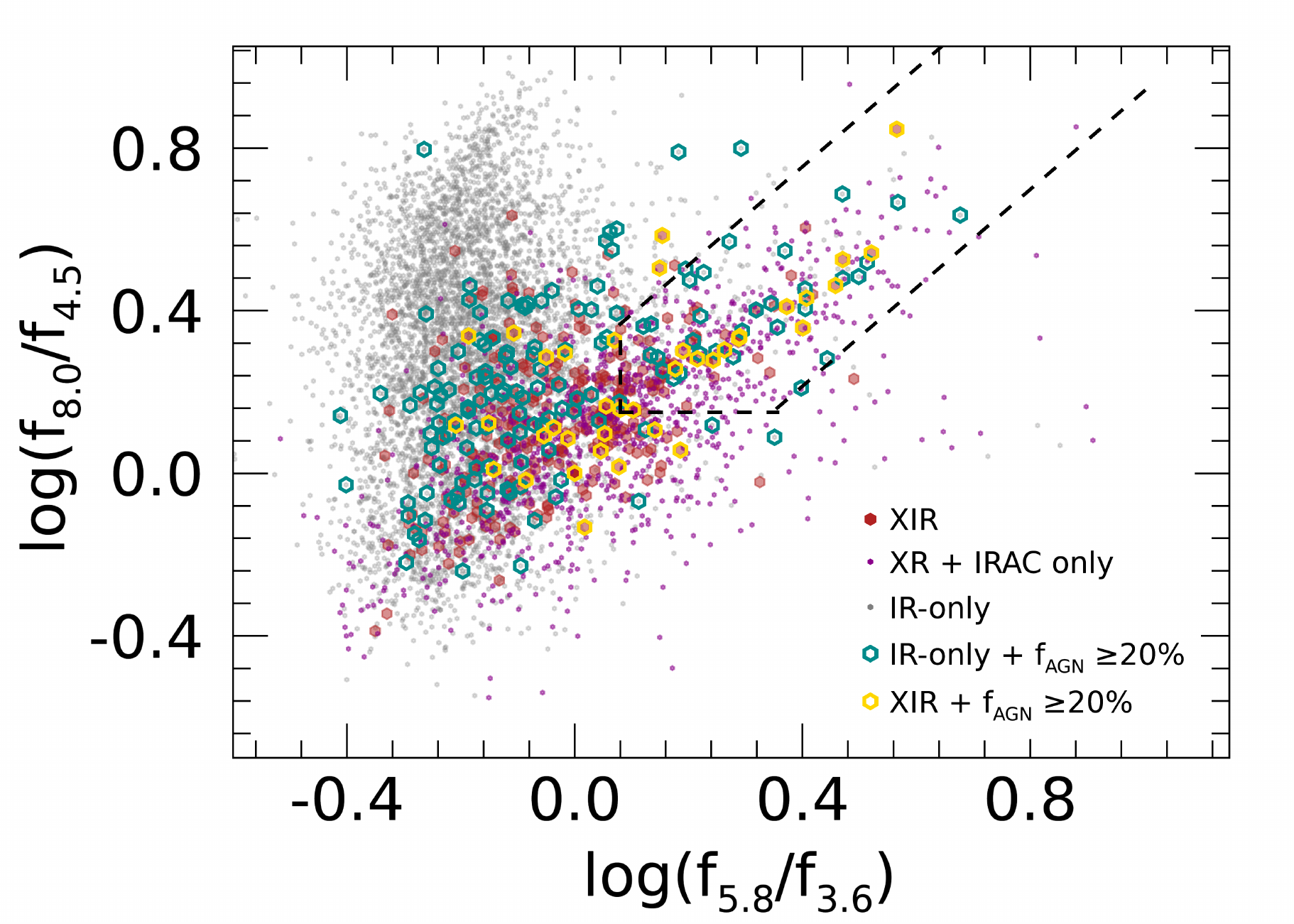}
&
\includegraphics[width=0.32\textwidth]{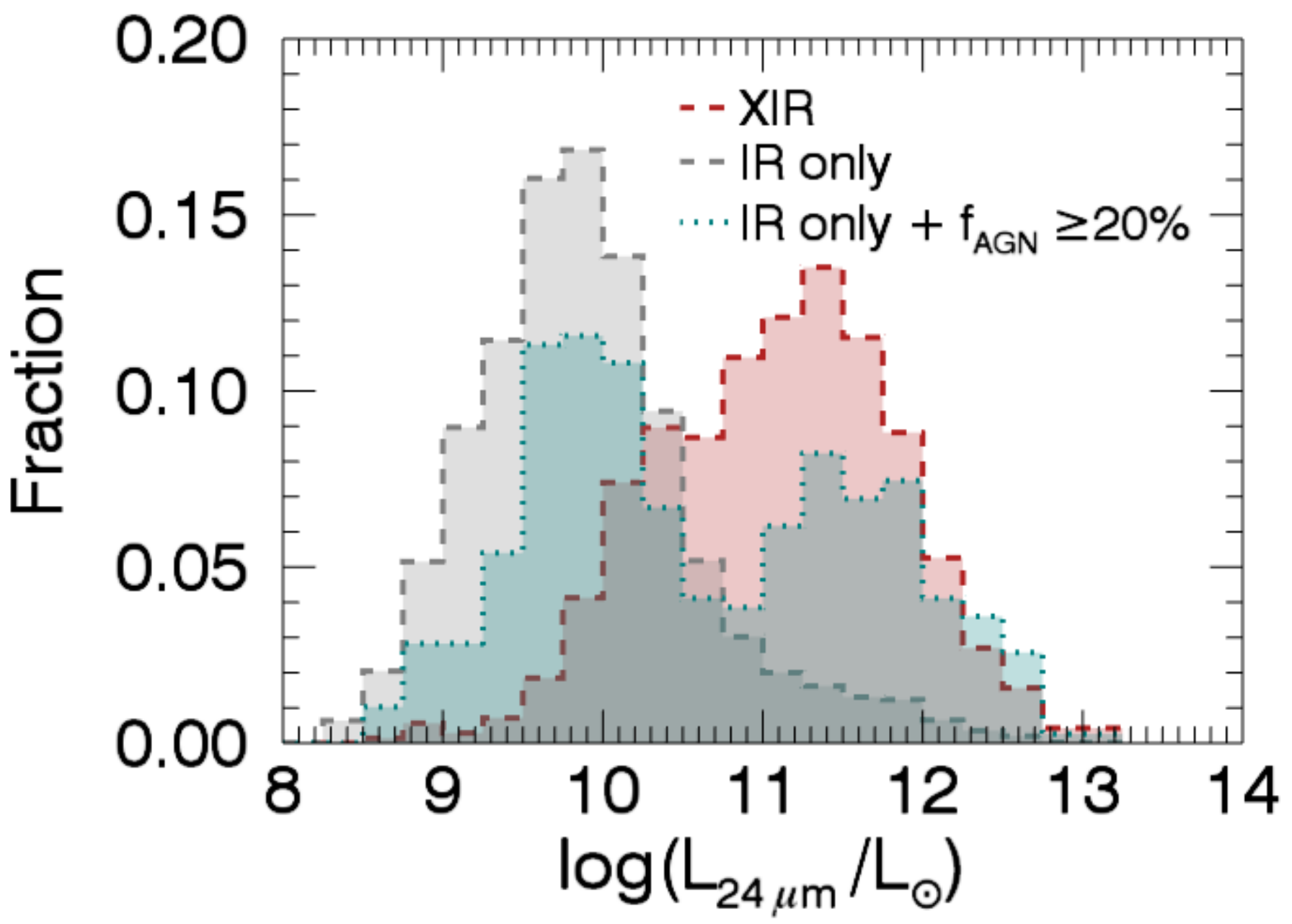}
&
\includegraphics[width=0.32\textwidth]{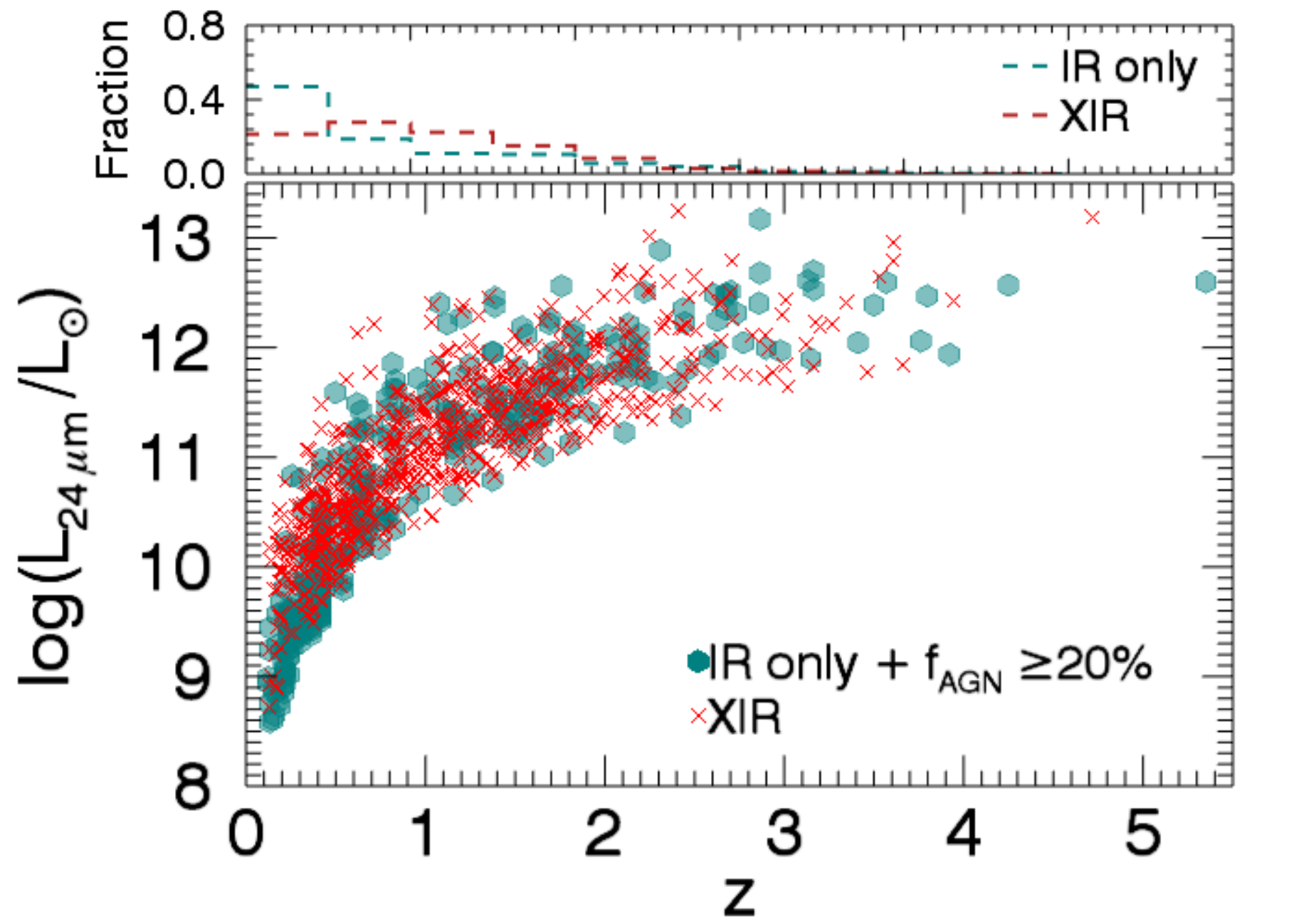}
\end{tabular}
\caption{\textit{Left}: IRAC color-color space used to identify luminous AGN in \citet{donley12}. Grey points represent the IR-only sources with spec-zs, large red dots mark sources from our main sample of XIR sources, and purple points are additional X-ray sources with IRAC counterparts, but no mid or far-IR counterparts. The dashed black lines carve out the region belonging to luminous AGN, with minimal contamination from high redshift star-forming galaxies. We note that 15\% of the luminous AGN in the IR only sample also have estimated AGN IR contributions $\ge 20\%$ (teal circles), while the same is true for 25\% of our XIR sample (gold circles). Generally, sources with AGN IR contributions $\ge 20\%$ are dispersed throughout this IRAC color space, indicating that SED decomposition does not lend itself to luminous AGN identification. \textit{Middle}: L$_{24\,\mu m}$ distribution for the 703 X-ray detected sample (red), the ~5k IR-only detected sample (grey), and the 389 IR-only sources with $\geq 20\%$ of IR SED emissions coming from AGN processes (teal). \textit{Right}: L$_{24\,\mu m}$ vs. redshift distribution for both the 389 IR-only sample with significant IR AGN contribution (teal) and the X-ray detected sample (red). Top histogram represents the fraction of sources from each sample in redshift bins of size 0.5.}
\label{zv24}
\end{figure*}

This study is based on the 703 X-ray sources in the Bo\"{o}tes field that have intensive multiband data to fit their individual spectral energy distributions (see Table \ref{fluxtable} for exact counts per band). Generating individual SEDs allows us to avoid the restrictions and uncertainties related to stacking and gives us the freedom to disentangle AGN and host galaxy radiation components for each respective source. Using \textsc{sed3fit} \citep{berta}, a multi-component SED fitting tool, we decompose each galaxy\textquoteright s emissions in the infrared spectrum and use the appropriate rest-frame, infrared luminosity integrated from 8\,$\mu$m to 1000\,$\mu$m as an indicator of host galaxy star formation rate. \textsc{sed3fit} is based off of the \citet{dacunha} \textsc{magphys} code and employs a combination of three galaxy radiation processes: stellar emission, warm and cold dust emission from star formation regions, and AGN emission. SED templates are fitted to measured fluxes first using the stellar and star forming components only, then AGN templates are varied to fill in photometric gaps and further reduce the $\chi^{2}$. We use the ten AGN templates provided with \textsc{SED3FIT}, which were selected to cover the wide range of AGN found in the full \citet{Fritz+2006} library. These ten templates span Type 1, intermediate, and Type 2 AGN, with a variety of optical depths ranging from 0.1-6, as viewed face on or edge on. All ten of the templates have a fixed torus opening angle of $\Theta = 100^{\circ}$, corresponding to an intrinsic covering factor of ~75\% (see Section \ref{dustcovering} for details on covering factors). Each AGN template can be broken down into three components: dust scattering emission, dust thermal emission and nuclear accretion disk emission. The former two AGN components combined are attributed to the warm, dusty clumpy structure that surrounds the SMBH and accretion disk. See Figure \ref{seds} for two example spectral energy distributions generated from our sample (left: star formation dominant, right: AGN emission dominant). 

	To correct for contaminating AGN radiation, we subtract the dusty torus and accretion disk emission from the total SED of a source. The resulting infrared luminosity is attributed to star formation and is hereby represented as L$_\text{IR}^\text{SF}$, while the subtracted infrared AGN luminosity is referred to as L$_\text{IR}^\text{AGN}$; Figure \ref{hist} (top) shows our resulting population distribution of infrared luminosities attributed to star-formation processes. This procedure applies to 98\% of our sample as 13 sources are not fitted with an IR AGN component by \textsc{SED3FIT}. The physical characteristics derived from this procedure will be available for all 703 sources on Vizier.\footnote{http://vizier.cfa.harvard.edu/}

	Out of the remaining 2.6k XBo\"{o}tes sources not used in our XIR sample, we also find 425 X-ray AGN with spectroscopic redshifts but no \textit{Spitzer} MIPS and \textit{Herschel} counterparts (marked as purple circles in Figure \ref{zvslx}) with a similar X-ray and redshift distribution as our main sample -- dubbed the non-IR sample (although some of these sources have IRAC detections; see next paragraph). We compare these non-IR AGN plus a sample of 6,583 IR-only galaxies with spectroscopic redshifts to our main sample in section \ref{results}. The IR-only galaxies have both a MIPS 24\,$\mu$m and at least one \textit{Herschel} far-IR detection, but no X-ray detection. For the non-IR AGN, we use the \textit{Herschel} SPIRE 250\,$\mu$m 5$\sigma$ limiting flux in the deeper region of the HerMES survey as a generous upper limit on star formation luminosity. Out of the IR-only galaxy sample, 99\% of sources have an optical counterpart and 91\% have an IRAC detection. We ran IR-only photometric data through \textsc{sed3fit} and found only 72\%($\sim$5k) of the $\sim$6.6k galaxies are fitted with an AGN component.

	For additional context, we briefly explore the additional two sample populations (6.6k IR-only galaxies and 425 non-IR AGN) in IRAC color-color space in Figure \ref{zv24} (left). Nearly 40\% of the non-IR AGN (small purple dots) and 92\% of the IR-only galaxies (grey points) have sufficient (3$\sigma$) detections in all four IRAC bands; the same is true for 46\% of our main XIR AGN sample (large red dots). In the \citet{donley12} IRAC color criteria for identifying luminous AGN (L$\mathrm{_{2-10 keV} \ge 10^{44} ergs\, s^{-1}}$; wedge outlined by dashed black lines), 60 of 327 XIR sources with detections in all four IRAC bands are categorized as luminous AGN with a median L$\mathrm{_{2-10 keV} \sim 5.6 \times 10^{43} ergs\,s^{-1}}$; only 23 of the 58 XIR sources with L$\mathrm{_{2-10 keV} \ge 10^{44}\,ergs \,s^{-1}}$ and IRAC detections are categorized as luminous AGN through the IRAC criteria, which is nearly equivalent to the X-ray luminous AGN recovery rate found in \citet{donley12} (38\%). This shows that, by using X-ray selection criteria, we're probing a larger population of the most powerful AGN. However, we must note that some powerful AGN are heavily obscured and therefore less X-ray bright ($30-60\%$, see Section \ref{dustcovering}); we caution readers to consider this selection effect throughout this work. 

	In the same space, 7\% of the non-IR AGN are categorized as luminous AGN with a median L$\mathrm{_{2-10 keV} \sim 2.3 \times 10^{44}\, ergs\,s^{-1}}$ and a recovery rate of 32\% for all X-ray luminous AGN in the non-IR sample; and out of the 5.6k IR only sources with spec-zs and sufficient IRAC detections, only 2\% (N\,$=128$) of sources are deemed luminous AGN (but members of this sample do not have any bona fied X-ray detections, so we cannot determine the recovery rate).

\begin{figure*}[!t]
\begin{tabular}{ll}
\includegraphics[width=0.5\textwidth]{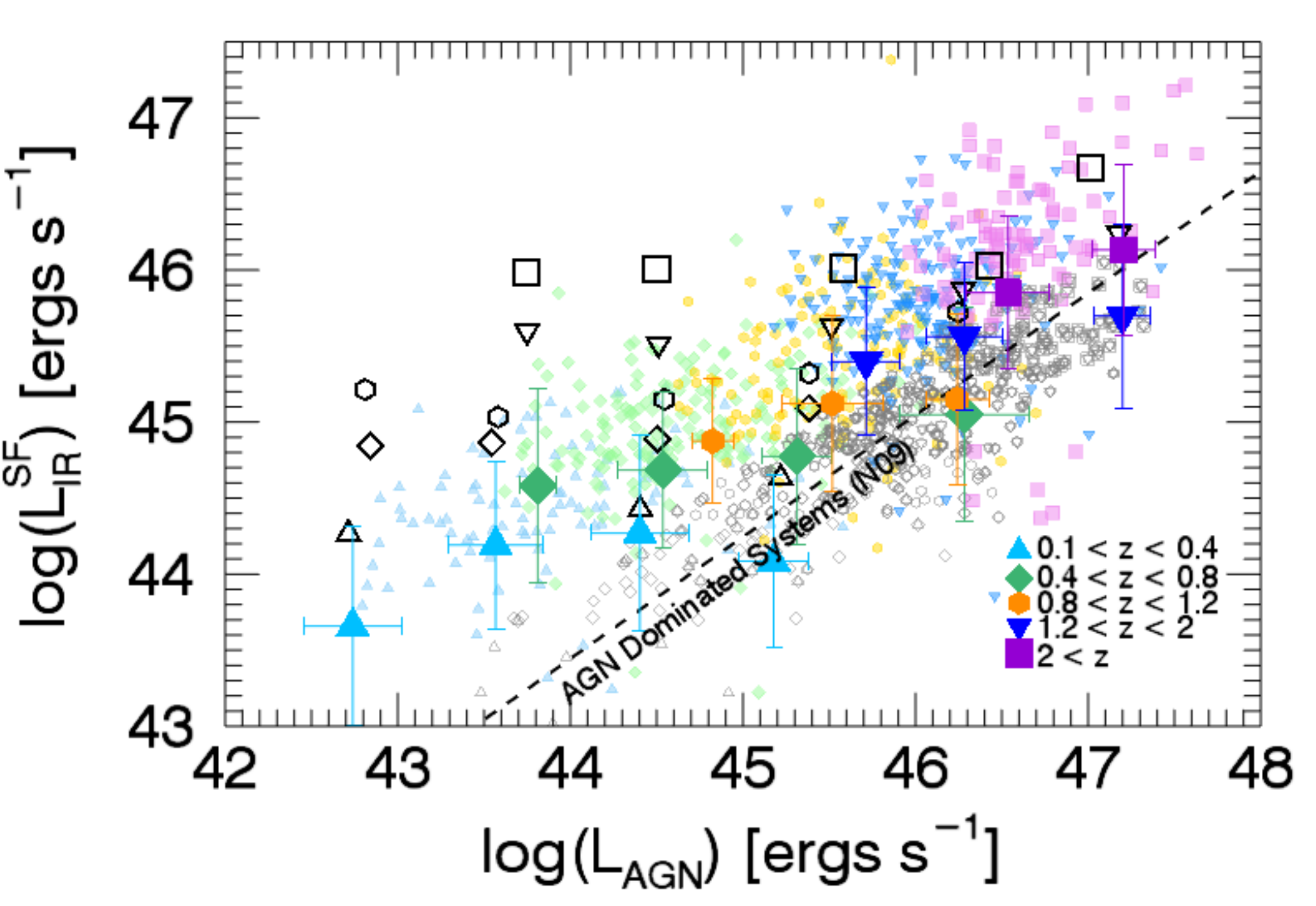}
&
\includegraphics[width=0.5\textwidth]{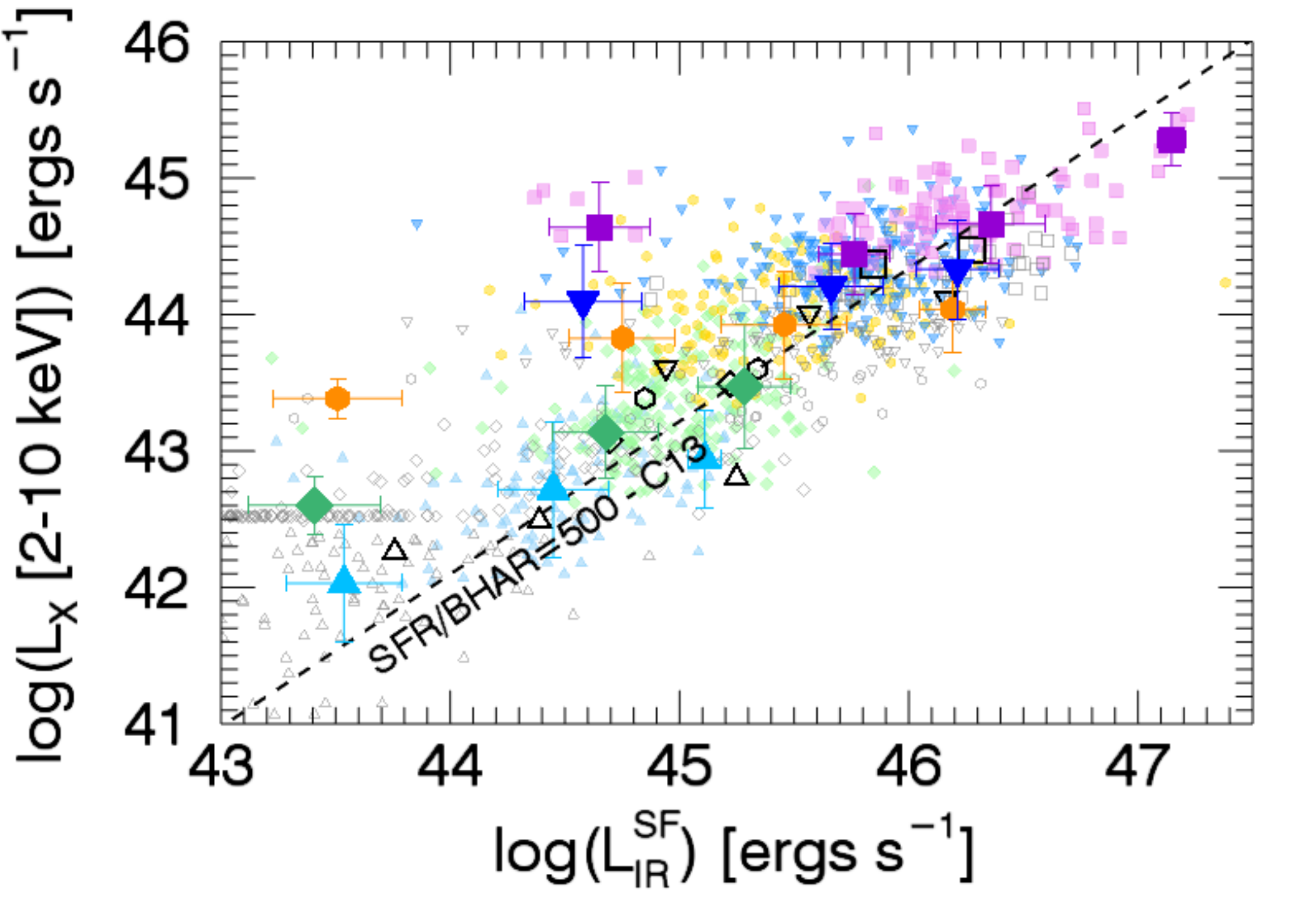}
\end{tabular}
\caption{\textit{Left}: Distribution of AGN bolometric luminosity ($\propto$\,L$_\text{X}$) versus L$\mathrm{_{IR}^{SF}}$. The lighter, smaller points are individual AGN. The small, empty gray symbols are X-ray AGN with L$\mathrm{_{IR}^{SF}}$ upper limits determined by the HerMES \textit{Herschel} SPIRE 250\,$\mu$m flux limit. The larger, bolder, filled in points are average log(L$\mathrm{_{IR}^{SF}}$) in bins of log(L$_\text{AGN}$) showing both the IR detected (colorful) and IR non-detected (empty grey) X-ray sources. Error bars represent the 1$\sigma$ dispersion of each bin. Note the star forming luminosity for the most powerful AGN in the $0.4 < z <0.8$ z-bin lies directly under the corresponding average star forming luminosity for the most powerful AGN in the $0.8<z<1.2$ z-bin. The black dashed line represents the relationship found in N09 where objects below the line have infrared luminosities  dominated by AGN activity. Black empty symbols are results from \citet{Lanzuisi}. \textit{Right}: Average log(L$_\text{X}$) in bins of log(L$\mathrm{_{IR}^{SF}}$) compared to results from C13 \citep{chen+2013}. The C13 sample is represented by the black empty shapes. The dashed line is the constant proportional relationship between star formation rate (SFR) and black hole accretion rate (BHAR) found in C13. Colors, symbols and error bars are calculated in the same fashion as in the left figure, where the empty gray points denote the IR only detected sources with an estimated IR AGN fraction $\geq 20\%$ with X-ray upper limits defined by the XBo$\ddot{o}$tes survey flux limit. }
\label{avgbins}
\end{figure*}

\begin{figure}
\includegraphics[width=0.5\textwidth]{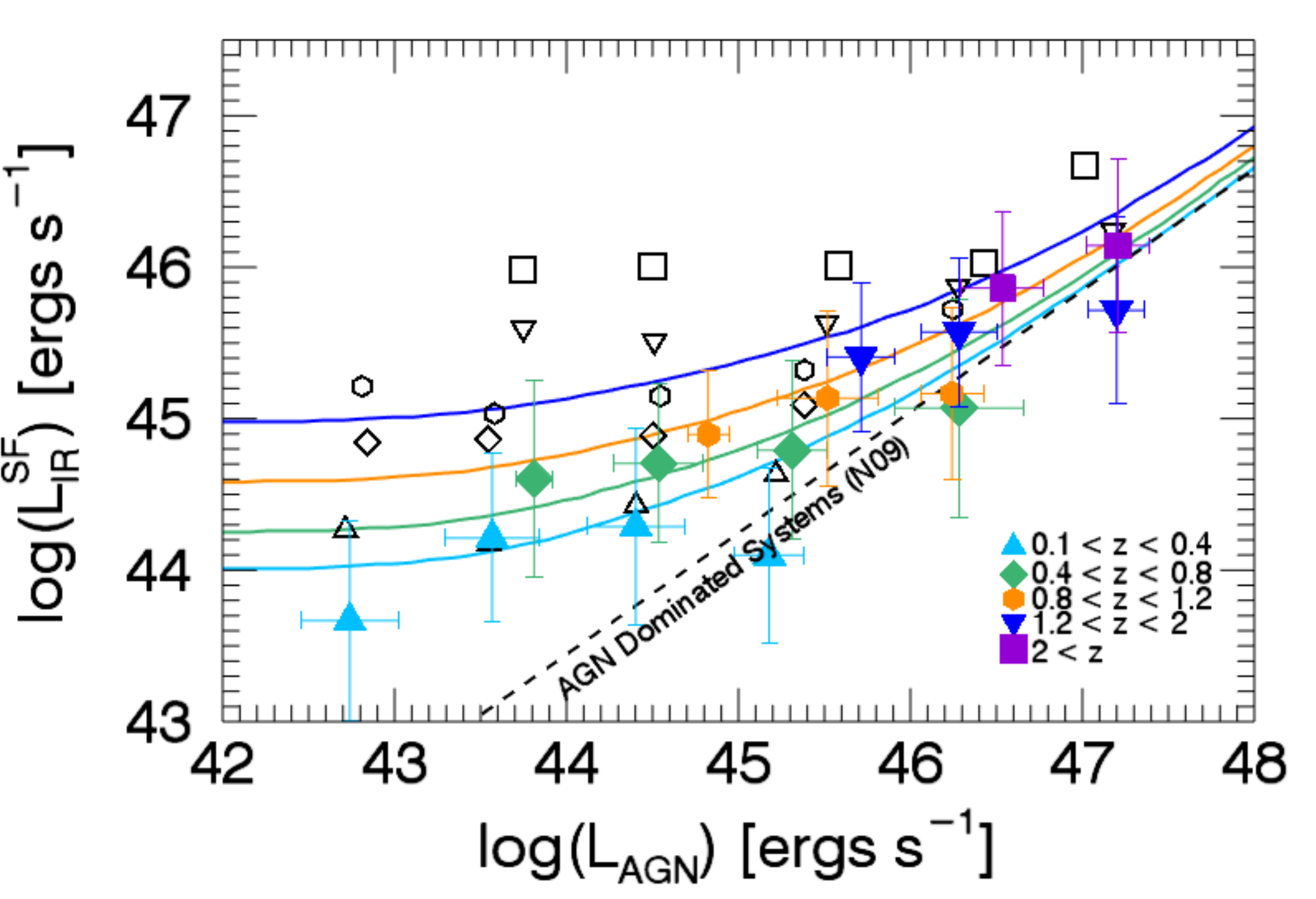}
\caption{Distribution of AGN bolometric luminosity versus L$\mathrm{_{IR}^{SF}}$ with the black dashed line as defined in Figure \ref{avgbins} (left). The solid lines have colors corresponding to redshift ranges and are the extrapolated trends from the \citet{hickox2014} simple model incorporating short-term AGN variability, long-term evolving star formation rates and a universal constant of proportion between SFRs and black hole accretion rates. }
\label{h14}
\end{figure}

\section{Results and Discussion} \label{results}
\subsection{Average L$_{IR}^\text{SF}$ vs. Average L$\mathrm{_{AGN}}$} \label{bolanal}

We translate X-ray flux to bolometric AGN luminosity, L$_\text{AGN}$, using the equation in \citet{Rosario+2012} (R12 hereafter) derived from \citet{maiolino} and \citet{NT} for spectroscopically confirmed Type 1 (unobscured) AGN:

\begin{equation} \label{agneq}
\mathrm{log\,L_{AGN} = \frac{log\,L_{X}-11.78}{0.721} + 0.845 }
\end{equation}

\noindent where L$_\text{X}$ is the 2-10 keV band X-ray luminosity. We average infrared contribution from star forming processes in bins of L$_\text{AGN}$, with respect to each redshift interval, and do the same separately for the additional 425 X-ray sources with spectroscopic redshifts but no IR counterparts. We show these results in Figure \ref{avgbins}(left); the dashed line represents the relationship found in \citet{n09} (N09, hereafter) for local, low luminosity AGN-dominated systems where L$_\text{AGN}$ is much larger than L$_\text{IR}$. Nearly 50\% of X-ray only detected sources fall into the AGN-dominated section, compared to only $\sim 5\%$ of individual X-ray and IR detected sources, substantiating the selection of AGN embedded within star-forming galaxies in this analysis and demonstrating the dominance of star-formation driven modes in IR luminosities of {\it Herschel} detected dusty galaxies. This trend is corroborated in several recent works using IR-bright X-ray selected AGN (e.g. R12, L17, \citealt{dai2015}), indicating that the power law correlation from N09 is valid when extended to higher luminosities and high-z AGN.

Our low-z ($z \lesssim 1$) sample successfully reflects those of other published results with low luminosity AGN (L$_\text{AGN} < 10^{45}\,$erg\,s$^{-1}$) showing a flat or uncorrelated relationship between AGN activity and star formation. The higher luminosity AGN in the low-z bins appear to trend in a more positive linear fashion that approaches the N09 relationship. The stronger, positive relationship is most noticeable in the $ 0.4 < z < 0.8$ bin where the most powerful AGN, while few in number (N=6), are embedded in star-forming galaxies nearly just as bursty as the brightest AGN in the $ 0.8 < z < 1.2$ bin. These results also appear in L17 and R12, but conflict with the flat, nonexistent relationships found in \citet{stanley} and \citet{dai2015}.

\citet{hickox2014} and \citet{volonteri15} developed models that match similar observational results as seen in L17, R12, \citet{chen+2013} and \citet{azadi15}. In Figure \ref{h14}, we overlay the \citet{hickox2014} model curves and see general agreement with the results for our $z\sim1$ less powerful active galaxies (L$_\text{AGN} < 10^{45}$\,ergs\,s$^{-1}$), but the model over estimates star forming luminosity for the more powerful AGN (L$_\text{AGN} > 10^{45}$\,ergs\,s$^{-1}$) in each redshift range. To create the model, \citet{hickox2014} generated a sample of galaxies (up to $z=2$) in which all star-forming galaxies host an AGN during their lifetime, and then incorporated a constant of proportion between SFR and black hole accretion rate over long time scales (log(SFR/BHAR) $= 3.6$ \citep{chen+2013, dai2015}) and assigned short time scale variabilities in AGN accretion processes (and therefore, luminosity). Generally, the model successfully produces the observed findings when averaging star formation activity in bins of AGN activity, along with the trends observed in literature when averaging AGN activity in bins of star-formation activity, as analyzed in the following section.

\subsection{Average L$_\text{X}$ vs. Average L$\mathrm{_{IR}^{SF}}$} \label{chenanalysis}
Recent simulations and observations reveal that AGN accretion (and therefore luminosity) can be highly variable on short timescales -- e.g. on the order of 1-2 magnitudes within 0.1-1 Myr \citep[e.g.][]{dimat05, hickox2014} -- whereas star formation processes change at a slower rate over longer timescales. To uncover the relationship between AGN processes and host galaxy star formation rates, it might be more appropriate to average AGN activity (the more rapidly changing variable) based on L$_\text{IR}^\text{SF}$ (the more stable variable). 

Following the analysis in L17 and \citet{chen+2013} (C13, hereafter), we reversed data dependency by averaging log(L$_\text{X}$) in bins of log(L$\mathrm{_{IR}^{SF}}$). We include 389 IR only sources with an AGN IR contribution that is $\geq 20\%$ of the total IR SED. These IR-only sources have both a MIPS 24\,$\mu$m and at least one \textit{Herschel} far-IR detection, but no X-ray detection (see Figure \ref{zv24} for 24\,$\mu$m population distribution). We take the ratio of IR AGN luminosity to total IR luminosity from the resulting SED and place a cut at $\geq 20\%$ to capture the sources with the highest likelihood of hosting an AGN \citep{ciesla}. For these objects, we use the XBo$\ddot{o}$tes survey flux limit as an upper limit for X-ray luminosity. Results are shown in Figure \ref{avgbins} (right) with L17 results overlaid. Error bars represent the 1$\sigma$ dispersion of the mean X-ray luminosity in each respective bin. The dashed line represents the constant ratio between black hole accretion rate (BHAR, proportional to X-ray luminosity) and star formation rate found in C13 for 34 X-ray detected AGN at $z=0.25-0.8$. 

We find our results to be in good agreement with the C13 SFR/BHAR ratio. The low z-bins ($z \lesssim 1$) have the strongest positive slope between the same L$\mathrm{_{IR}^{SF}}$ intervals studied in C13, which is expected as C13 analyzed data from the same Bo\"{o}tes \textit{Chandra}, \textit{Herschel} and \textit{Spitzer} observations used in this paper. While AGN still hover near the SFR/BHAR ratio in the earlier $z > 0.8$ Universe, there is no significantly strong upward trend as L$\mathrm{_{IR}^{SF}}$ increases for any z-bin, and the nearly $\sim$0.5 dex increase exhibited within the $z > 2$ sample for the highest range of star formation activity has a very small sample size and is therefore unreliable. 

Note that these observations are limited to the depths of the 24\,$\mu$m survey; an object at $z\sim1$ with a 24\,$\mu$m luminosity of $L_{24\,\mu m} = 10^{44}$\,ergs\,s$^{-1}$ is pushing the survey observational limits and might be undetected. This means that the weakest star formation bins in this analysis may be lacking contributions from some fainter, intermediate redshift galaxies and AGN. Conversely, some powerful AGN are heavily obscured by high column densities of dust and gas. In fact, studies have shown that 90\% galaxies with high 24\,$\mu$m to optical flux ratios have IR and X-ray signatures indicating the presence of heavily obscured AGN \citep{fiore08,treister09}. These AGN are expected to have intrinsic X-ray luminosities in excess of $10^{44}$\,ergs\,s$^{-1}$, at $z \sim 1-2$, which could drive the more star-forming L$\mathrm{_{IR}^{SF}}$ bins further upward and into stronger agreement with the C13 trend.

The observed differences in correlation between the two averaging methods are likely due to the inherent rate of variation between the two physical processes, with star formation being the more stable measurement and AGN accretion being the more variable measurement. These differences in correlation methods were also confirmed by \citet{dai2015} for similar samples of X-ray selected AGN. \citet{lapi2014} found similar results when exploring the observational phenomena of the co-evolutionary relationship between AGN and host galaxies at high redshifts ($z \gtrsim 1.5$) using a semi analytical model. Combining observational data on AGN in star-forming galaxies with high-$z$ AGN luminosity functions and host-galaxy stellar luminosity functions, the model shows galaxy SFRs that remain relatively constant over a long period of time then suddenly undergo a rapid decrease in star formation when the SMBH is triggered into an active phase. The model also predicts that as the supermassive black hole grows, a fraction of the cold interstellar gas and dust within the spiral arms of a galaxy is drawn towards the nucleus to help form and grow the dusty torus. The AGN will feed off this reservoir and the most powerful AGN will have feedback processes that strip away some of the remaining cold gas and dust, further suppressing star formation processes and eventually slowing its own growth as well. Observations at various epochs within the model easily reproduce both of the trends shown in Figure \ref{avgbins} and, when combined with the publications and findings discussed in section \ref{bolanal}, indicate that a more detailed study on the relationship between short term AGN variability and host galaxy cold gas and dust properties is necessary to arrive at any definitive conclusions.

\begin{figure*}[!ht]
\begin{tabular}{ll}
\includegraphics[width=0.5\textwidth]{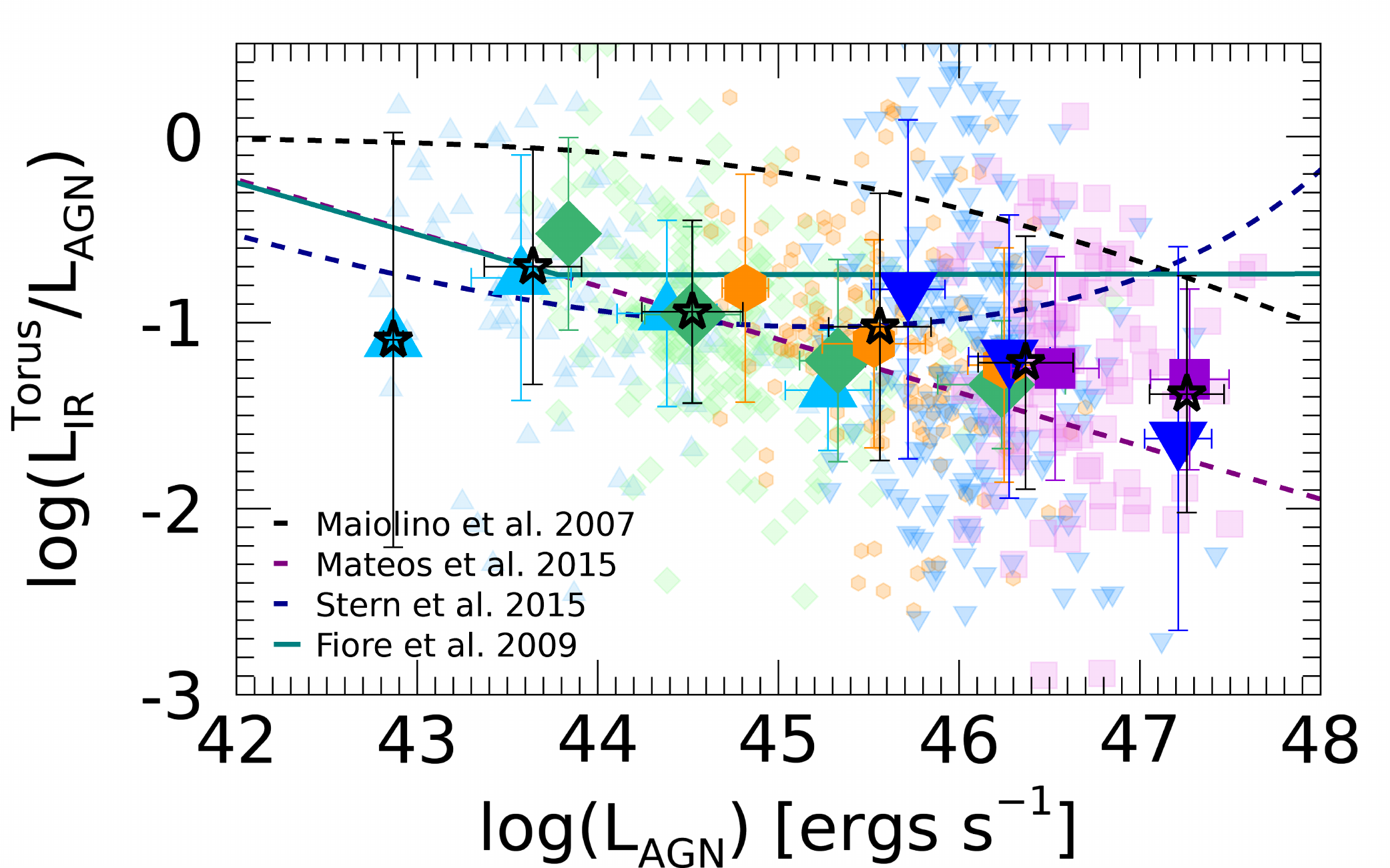}
&
\includegraphics[width=0.5\textwidth]{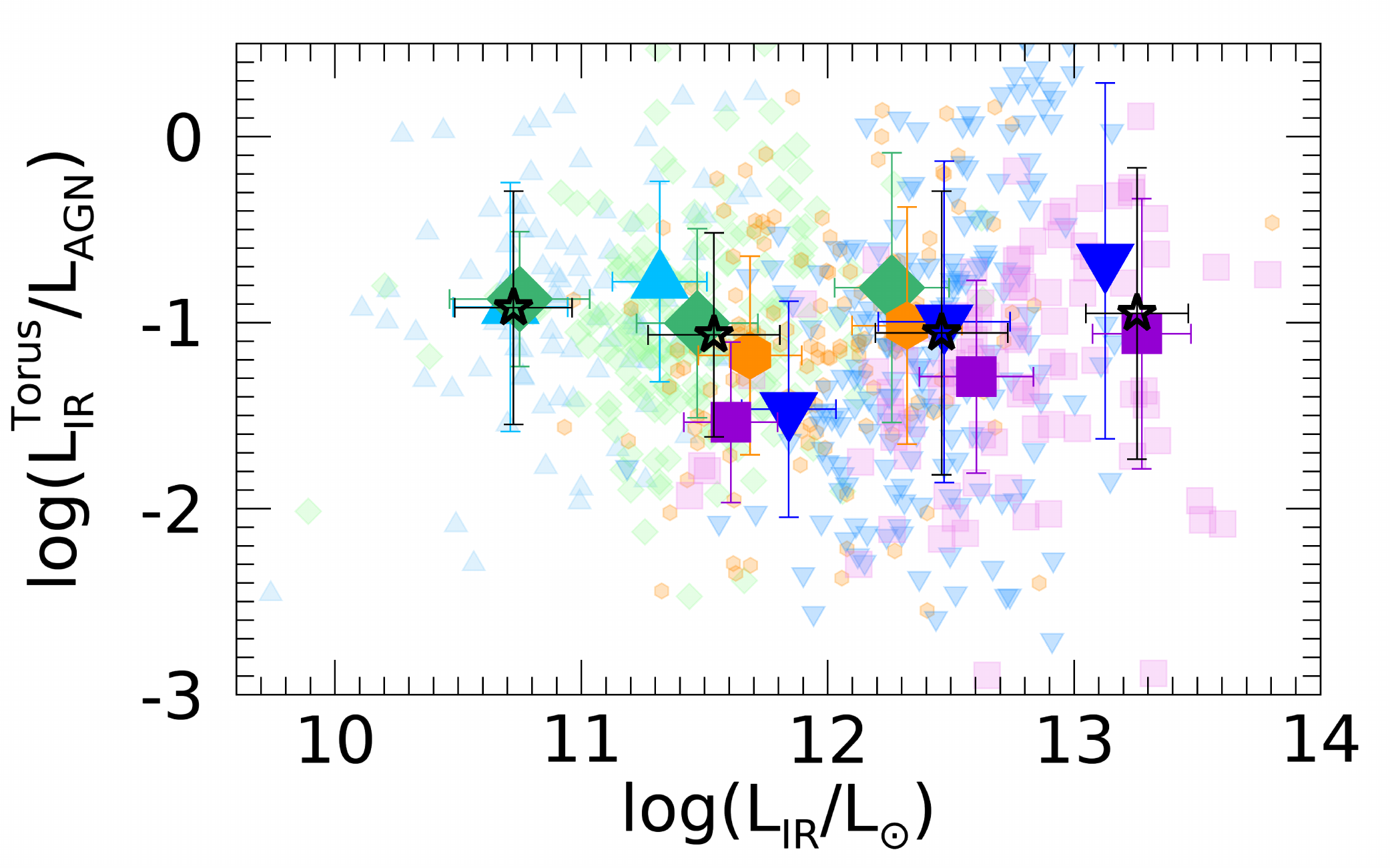}
\end{tabular}
\caption{\textit{Left}: Covering factor versus bolometric AGN luminosity. Averages for the X-ray detected sample are computed in bins of L$\mathrm{_{AGN}}$ and in respective redshift ranges. We also computed averages for the entire sample, irrespective of redshift range, as indicated by the empty black stars. The black dashed line represents the fraction of obscured AGN as a function of bolometric AGN luminosity found by \citet{maiolino}. The purple dashed line, navy dashed line, and turquoise solid line correspond to mid-IR/L$\mathrm{_{AGN}}$ fractions found by translating the X-ray-to-6\,$\mu$m relationships derived in \citet{mateos15}, \citet{stern15}, and \citet{fiore}, respectively. \textit{Right}: Covering factor versus total infrared luminosity. Averages, colors and symbols are derived in the same fashion as the figure to the left. }
\label{CFs}
\end{figure*}

\begin{figure}
\includegraphics[width=0.5\textwidth]{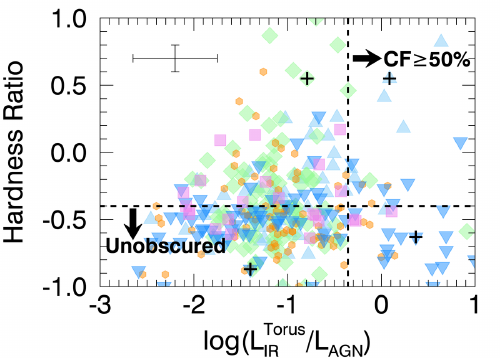}
\caption{Covering factor versus hardness ratio for the 330 XIR AGN with sufficient x-ray counts to determine hardness ratios. Colors and shapes are the same as those in Figure \ref{avgbins}. Average error bars are presented in the top left corner. To the right of the black dashed vertical line lies AGN with covering factors greater than 50\%. Below the black dashed horizontal line lies AGN with hardness ratios indicative of unobscured cores. SEDs for sources marked with crosses are in the appendix, providing examples of some of the more extreme and contradictory AGN in this sample.}
\label{hrs}
\end{figure}

\subsection{Dust Covering Factors} \label{dustcovering}
We can determine how dust obscured an accreting SMBH is by assessing the relationship between how much high energy radiation from accretion disk processes is observed (which therefore escapes the dusty torus), versus how much radiation is detected from the dusty torus itself. A commonly used dust covering factor (CF) proxy is the ratio of dusty torus emission, L$\mathrm{_{Tor}}$ (which dominates in the mid to far-IR), to bolometric AGN luminosity, L$\mathrm{_{AGN}}$ \citep[e.g.][]{maiolino, treister08, rr09}. To compute the dust covering factor for our sample, we use the bolometric AGN luminosities derived from Equation (2), and derive L$\mathrm{_{Tor}}$ from the dusty torus components in each source's respective AGN SED (i.e. we remove the infrared emission originating solely from the accretion disk from each AGN SED template for each source and keep only the dusty torus emission components). We caution that systematics from the fixed covering factor ($75\%$) in the AGN SEDs may produce biased estimates of dusty torus emission in this analysis (see Section \ref{samplesel}).

	We note that this proxy (CF\,$=\mathrm{L_{Tor}}/\mathrm{L_{AGN}}$) is used under the assumption that accretion disk emission and the resulting dusty torus emission are generally isotropic. However, the work of \citet{stalevski} shows that, when considering the anisotropy of these emission processes for Type 1 AGN with L$\mathrm{_{AGN}} \sim 10^{45}$\,ergs\,s$^{-1}$, this proxy can underestimate intrinsically low covering factors and overestimate high covering factors, while for Type 2 AGN of similar luminosities, this proxy always underestimates the true covering factor. We assess the impact of this assumption on our results at the end of this section.

\begin{figure*}[!t]
\begin{tabular}{ll}
\includegraphics[width=0.5\textwidth]{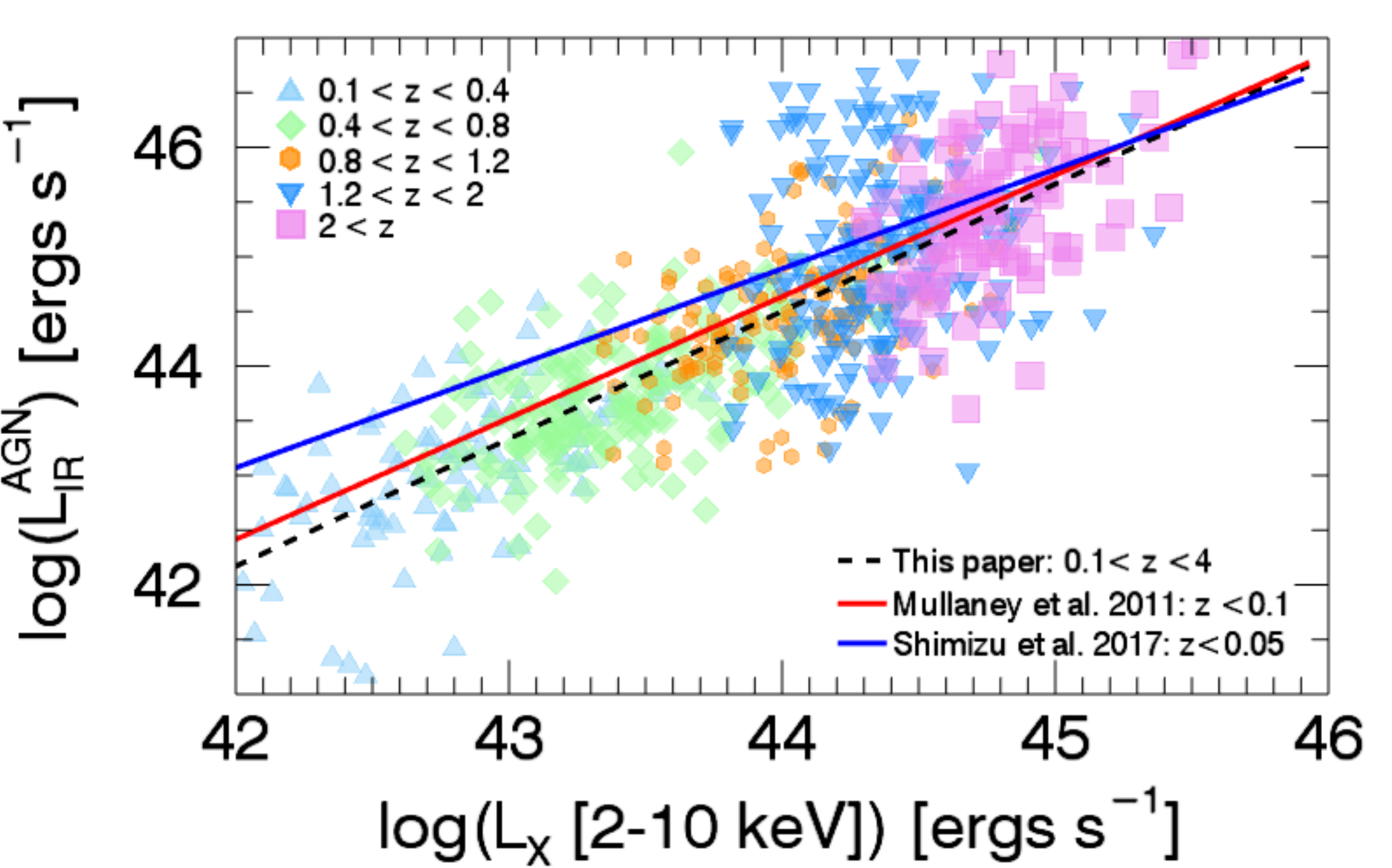}
&
\includegraphics[width=0.5\textwidth]{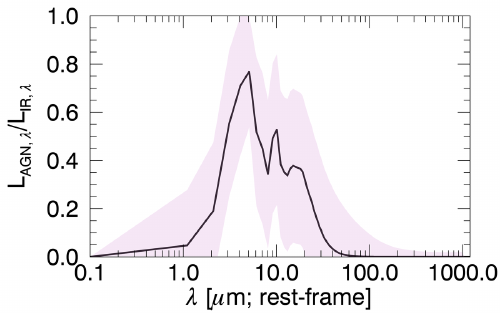}
\end{tabular}
\caption{ \textit{Left}: Infrared AGN luminosity as a function of L$\mathrm{_{X}}$ for the X-ray AGN. The black dashed line represents the linear relationship found in log space between the two AGN luminosities. The red and blue lines represent the relationship determined from generated average SEDs for local AGN by \citet{mullaney} and \citet{shimizu}, respectively. \textit{Right}: Composite $f\mathrm{_{AGN}}$ as a function of rest-frame wavelength using all 703 X-ray selected AGN. The black line is the median value at all wavelengths in bins of $\Delta\lambda = 1\,\mu$m and the 1$\sigma$ scatter for each $\Delta\lambda$ is indicated by the shaded pink region.}
\label{fagn}
\end{figure*}

\begin{figure}[h]
\includegraphics[width=0.48\textwidth]{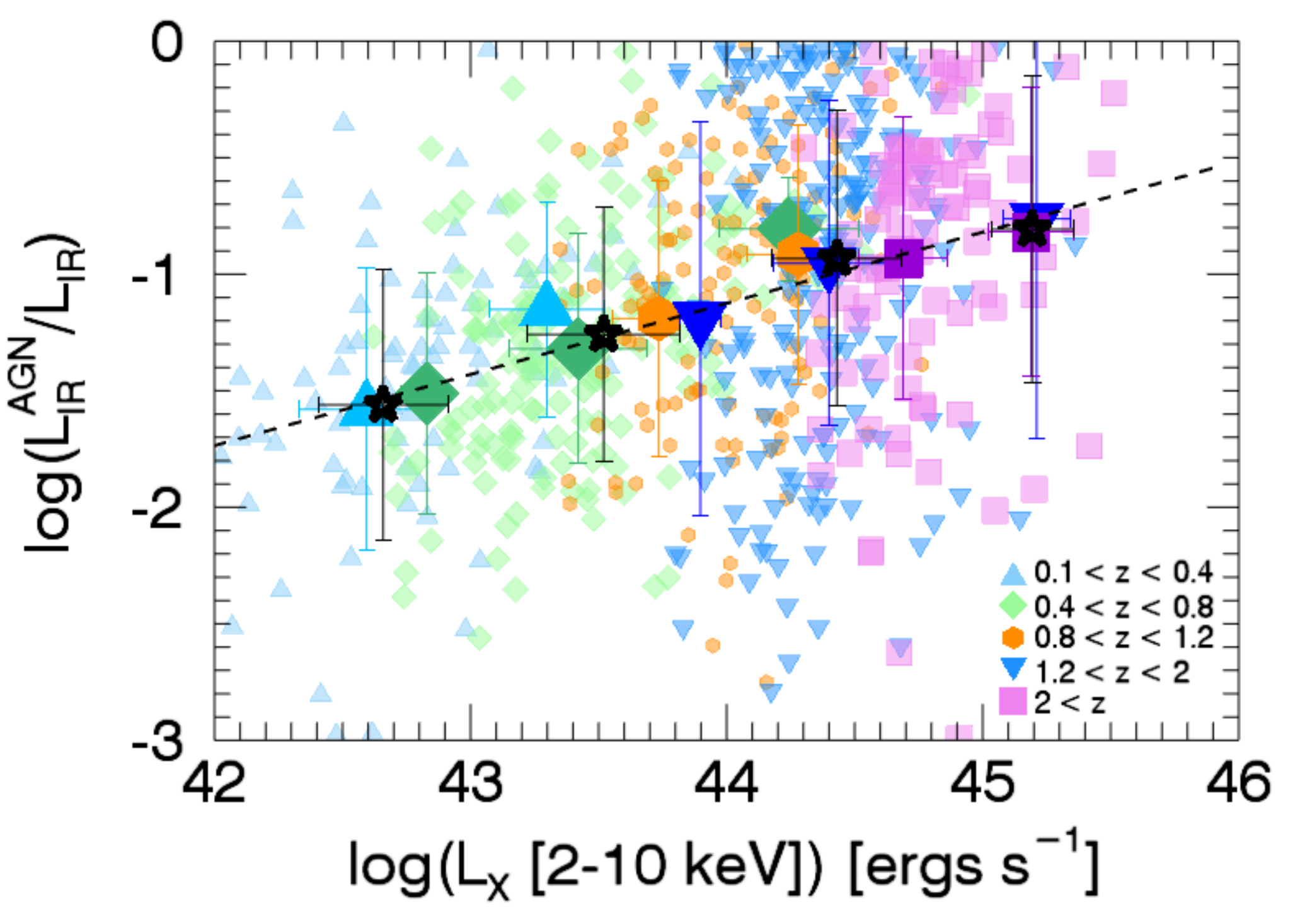}
\caption{ AGN fractions for the X-ray detected sample as a function of X-ray luminosity. Averages are computed in bins of L$_\text{X}$, respective of redshift range, with error bars representing the 1$\sigma$ dispersion of the mean. Black empty stars represent the averages across X-ray luminosity, regardless of age in the universe.}
\label{xratios}
\end{figure}

The average dust covering factor decreases with an increase in AGN activity for our X-ray detected AGN sample (Figure \ref{CFs}, left). This trend correlates nicely with the luminosity-dependent AGN unified model where dust covering factor is anti-correlated with bolometric luminosity, also known as the \textit{receding torus model} \citep{lawrence}. Taking the model implications a step further, it follows that the average covering factor within a sample of AGN corresponds directly to the fraction of Type 2 (obscured) AGN. In this work, we find an average CF of 33\% for the X-ray detected AGN. This average CF is similar but slightly lower than those found in literature: \citet{rr09} used \textit{Chandra} and/or \textit{Spitzer} data to determine CFs for 658 AGN and found an average dust covering factor of 40\%; \citet{mateos15} determines a spectroscopically confirmed Type 2 fraction of 43\% on a sample of 250 X-ray selected AGN with dust covering factors ranging from 20-50\% when averaged in bins of X-ray luminosity; \citet{lanzuisi09} found a higher Type 2 fraction at 55\% of mid-IR bright X-ray selected AGN, and \citet{hickox07} selected IR-AGN in the same field as this study and used spectroscopic and optical to mid-IR color distributions to determine a Type 2 fraction of 43\%. The observed luminosity-dependence agrees most with the trend found in \citet{mateos15} (shown as the purple dashed line in Figure \ref{CFs}, left), who also used multi-component SEDs to determine the AGN contribution to mid-IR luminosity. A newer study by \citet{mateos17} investigated the lack of one to one correlation between Type 2 fraction and average covering factor for their complete sample of optically classified X-ray AGN. They identify a missing population of X-ray obscured AGN and, when the high covering factors of these obscured AGN are accounted for, the population CF average grows to nearly 60\% with a less significant luminosity dependence relationship. It is possible that the CFs of heavily obscured AGN in the Bo$\ddot{o}$tes region would effectively raise the average CF across all redshift ranges and AGN luminosities to a similar value, but that analysis it out of scope for this work.

In Figure \ref{CFs}, right, we find an overall flat relationship between total infrared luminosity and covering factors for the X-ray selected sample, hovering at an average of $\sim10\%$ across all luminosities. While there appears to be some positive relationship for all redshift bins $z>0.4$ starting at log(L$\mathrm{_{IR}}$/L$_{\odot}$)$\approx 11.5$, the sample dispersion is large, spanning $\pm\sim50\%$ (or more) for each average data point within each redshift bin. Therefore, any observed positive correlation is weak and would require further investigation for verification. 
	
	We also recover trends that challenge the inclination-based unified model: there is no clear bimodal distribution for covering factors in the XIR AGN population; instead we see a distribution of covering factors that cover the entire possible range at significant percentages. To investigate, we further restrict our sample to the 330 XIR AGN with sufficient X-ray counts to determine hardness ratios (HRs; i.e. H$-$S / H$+$S, an indicator of AGN obscuration; e.g. \citet{green04}) and find the majority ($\sim57\%$) are unobscured with corresponding HRs\,$ \lesssim -0.5$ and an overall wide spread in covering factors averaging at $35\% \pm 1.03\%$ (see Figure \ref{hrs}). Concentrating only on the 187 XIR AGN with unobscured HRs, we find $11\%$ have CFs $\gtrsim 50\%$, indicating that a defining CF cut off limit between Type 1 and Type 2 AGN based on X-ray absorption is nonexistent. \citet{mateos16} found similar results using 227 spectroscopically confirmed and categorized X-ray AGN; while the different types of AGN had clearly different CF distributions (with type 2(1) peaking at high(low) covering factors) there was still a very strong overlap in CF distributions; roughly 20\% of Type 1 AGN had CFs $ > 0.5$ and 40\% of Type 2 AGN had CFs $< 0.5$. \citet{merloni} used optical photometry and/or spectra paired with hard X-ray data for $\sim$1300 AGN and found 31\% of the entire sample sits in a similar contradictory region where optical signatures point towards an unobscured nucleus while X-ray data indicates considerable gas and dust absorption, or vice versa with optical evidence for an obscured nuclear region and no absorption of soft X-rays. This work and the aforementioned suggest that Type 1 and Type 2 AGN may not be observationally distinct due to the line-of-sight inclination of the dusty torus but instead due to other physical accretion related mechanisms.
	
	Recently, \citet{ricci17} showed that the relationship between AGN luminosity and covering factor flattens out when dividing X-ray AGN into separate bins of Eddington ratios ($\lambda_{E}$; mass-normalized black hole accretion rate), indicating that the AGN line-of-sight obscuration is not the universal driver of covering factor distributions. Instead, $\lambda_{E}$ and CF maintain a steady positive correlation up until the sublimating Eddington limit for dusty gas particles, in which the CF sharply declines. These results point towards strength in radiation pressure from accretion activities being the main regulator of observed obscuration fractions, and that Type 1 and Type 2 AGN are actually physically different objects (as categorized by $\lambda_{E}$) that could be better unified within the context of black hole growth over time. Exploration of this relationship is out of scope for this analysis; we refer readers to \citet{beckmann,winter,ezhikode,lusso,lawrenceandelvis, mateos17} for further discussions that precede the \citet{ricci17} results.

\begin{figure*}[!t]
\begin{tabular}{ll}
\includegraphics[width=0.5\textwidth]{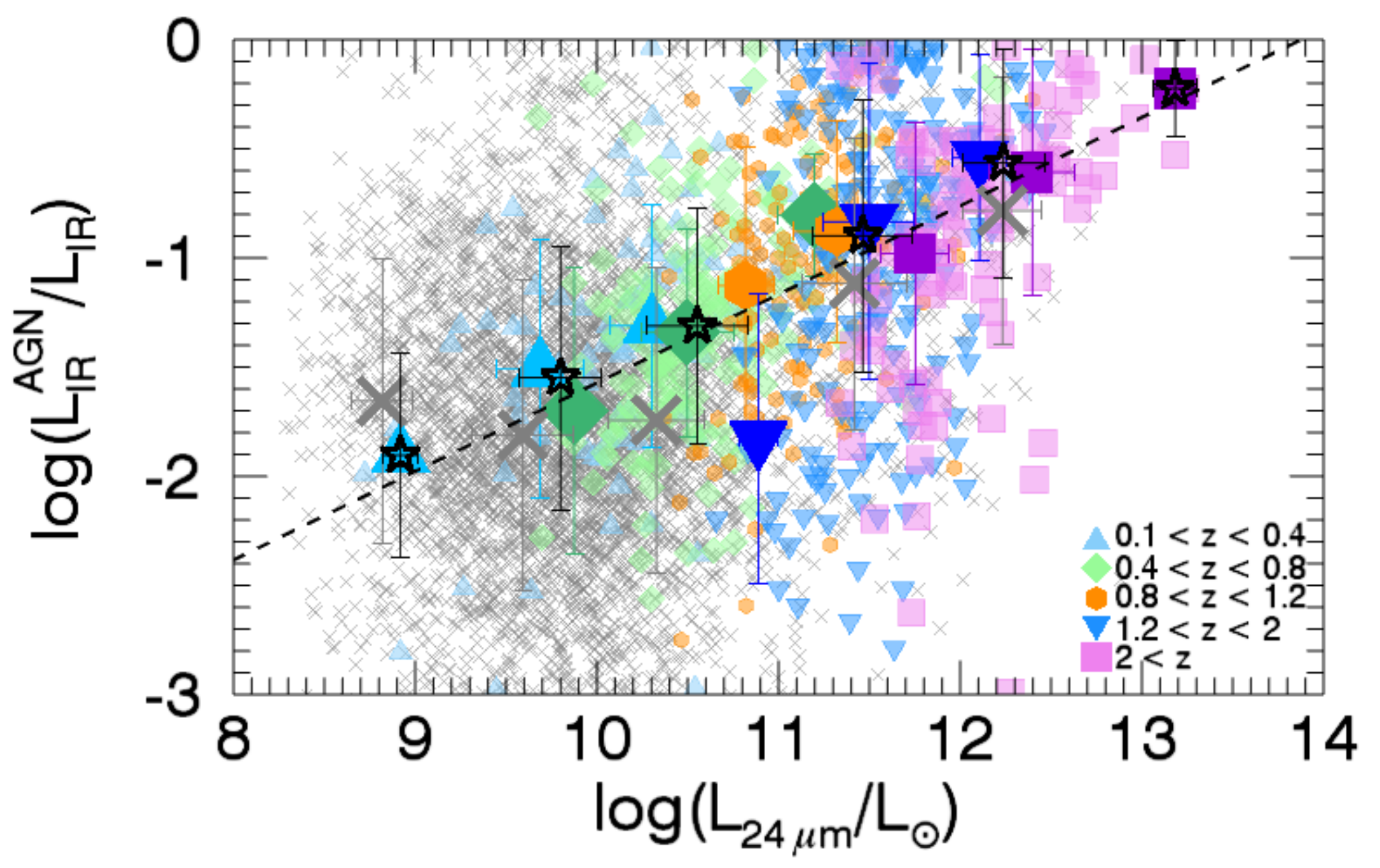}
&
\includegraphics[width=0.5\textwidth]{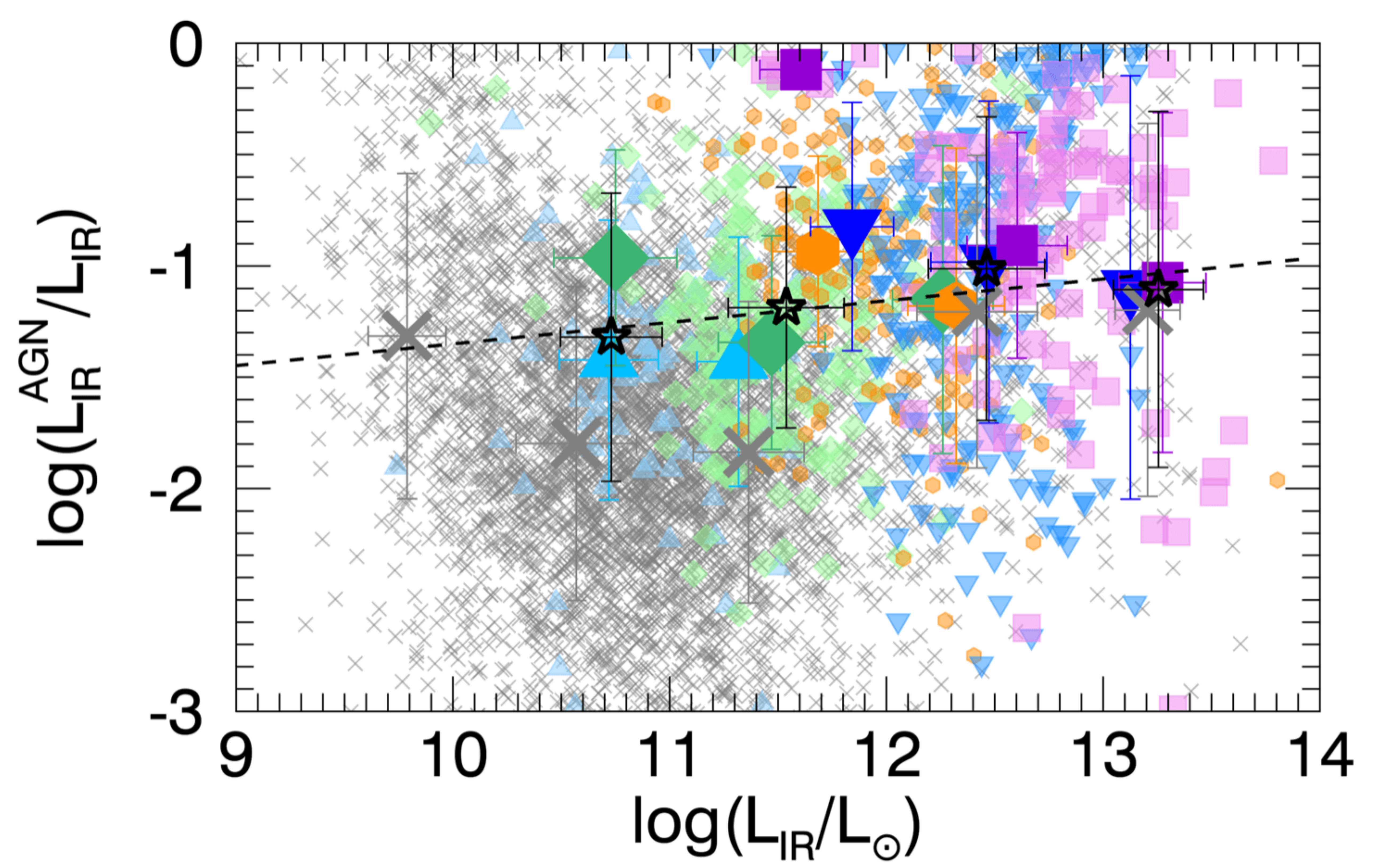}
\end{tabular}
\caption{These two figures represent the ratio of infrared AGN luminosity to host galaxy total infrared luminosity (i.e. AGN fraction) as a function of host galaxy luminosities. Colors and symbols are the same as in Figure \ref{CFs}, with grey x\textquoteright s denoting the individual $\sim6$k IR-only galaxies as defined in Section \ref{samplesel} and their averages in luminosity buckets of size 1\,dex are represented as large grey x\textquoteright s. \textit{Left}: The logarithm of the AGN fraction as a function of 24\,$\mu$m luminosity. The XIR sample shows a clear, but weak, correlation between 24\,$\mu$m luminosity and AGN fractions. \textit{Right}: The logarithm of the AGN fraction as a function of total infrared luminosity, following the same legend as in the figure on the left. There appears to be no clear relationship between total IR luminosity and AGN fraction, indicating a need for individual IR SED decomposition when estimating AGN fractions across IR luminosity space. }
\label{ratios}
\end{figure*}

	We explored how the \citet{stalevski} equation and coefficients (Equation 8 and Table 1, inside) for correcting isotropically-assumed dust covering factors affect our results by first identifying Type 1 and Type 2 AGN using the inclination angles used in the SED fitting procedure. The AGN SED fitting model used in this paper include two possible nuclear line of sight angles: 0 degrees (face-on aka Type 1 unobscured nucleus) or 90 degrees (edge-on aka Type 2 nucleus viewed through the disk). Based on this criteria, 61\% of our XIR sources are categorized as Type 1 AGN and the remainder are categorized as Type 2 AGN, which is consistent with the average CF derived earlier in this section. Interestingly, the majority of Type 2 AGN in this sample (77\%) have covering factors below $\le10\%$, while Type 1 AGN exhibit no general CF preference. Both AGN types have median AGN luminosities of L$\mathrm{_{AGN}} \sim 3\times10^{45}$ ergs\,s$^{-1}$. 

	We applied each set of coefficients corresponding to the three reported example optical depths ($\tau_{9.7 \mu \mathrm{m}} = 3,5,10$ in \citet{stalevski}) to the respective AGN types. The overall effect is strongest for AGN (of both types) with originally estimated CFs less than 20\%, which is nearly three quarters of the 703 AGN; for each set of coefficients, dust covering factors were increased to $\ge 20\%$, due to the lower limits assumed in \citet{stalevski}, with the average individual differences being $+33\%$ to the respective CFs. This effectively flattens out any trends seen in Figure \ref{CFs}, where the original average CF of 33\% is now a corrected average CF of 49\%. It is worth noting that these equations were originally derived for a luminous AGN with L$\mathrm{_{AGN}} \sim 10^{45}$ ergs\,s$^{-1}$; the third of our sample at lower AGN luminosities sees an average CF correction of $\sim+25\%$, while the remaining more powerful population has a ten percent higher average CF correction than that of the low luminosity AGN. Thus, due to the the underlying assumptions in covering factors and the wide range in AGN luminosities probed in this work, we are unfortunately limited from interpreting any further.

\subsection{AGN Contribution in the Infrared}
The wide infrared coverage in the Bo\"{o}tes region when paired with the multicomponent SED fitting model \textsc{sed3fit} is advantageous in effectively constraining intrinsic infrared AGN luminosities across a broad redshift range. This is useful to avoid situations of overestimating host galaxy properties (e.g. star formation rates) in cases with little IR photometry and/or possible indications of AGN activity. In the following, we explore the extracted infrared AGN luminosities in the $8-1000\,\mu$m range (L$\mathrm{_{AGN}^{IR}}$, hereafter) as a function of L$\mathrm{_{X}}$, as well as the fraction of total infrared luminosity attributed to AGN emissions (L$\mathrm{_{AGN}^{IR}}$/L$\mathrm{_{IR}}$ or $f\mathrm{_{AGN}}$, hereafter) as a function of L$\mathrm{_{X}}$, L$\mathrm{_{IR}}$, and L$_{24\,\mu m}$. 

There is a strong correlation between X-ray activity and total infrared AGN luminosity within our X-ray detected sample. This relationship is similar to the driving trend determined in \citet{mullaney}, even though a large portion of our sample contains galaxies with low AGN fractions ($f\mathrm{_{AGN}} < 10\%$) out to high redshifts. \citet{mullaney} modeled intrinsic infrared AGN SEDs for only 11 local ($z<0.1$) AGN with polycyclic aromatic hydrocarbon emission lines indicative of IR luminosities dominated by AGN ($f\mathrm{_{AGN}} > 90\%$). As seen in Figure \ref{fagn} (left), we derive a nearly equivalent relationship for AGN spanning a much larger redshift and AGN fraction range, suggesting that this relationship is universal. The black dashed line denotes our sample relationship, where

\begin{multline}
	\mathrm{log\,\bigg(\frac{L_{IR}^{AGN}}{10^{43} erg\,s^{-1}}\bigg)} = (0.33 \pm 0.06) \\ + \mathrm{(1.16 \pm 0.05)\,log\,\bigg(\frac{L_{X}}{10^{43} erg\,s^{-1}}\bigg)}
\end{multline} \label{xreqn}

\noindent with a strong, positive correlation coefficient of 0.78. The red line denotes the slope found in \citet{mullaney} ($1.11 \pm0.07$) and the blue line denotes the slightly weaker relationship found by \citet{shimizu} ($0.91 \pm0.06$) who analyzed a sample of 313 local X-ray selected AGN with \textit{Herschel} and \textit{WISE} detections; additionally, \citet{kirk17} found a more extreme relationship ($3.76\pm0.08$, not plotted) for 53 $z\sim1-2$ composite galaxies in the GOODS-S field with \textit{Spitzer} and \textit{Herschel} detections. Our work provides the first $0.1<z<4$ pure AGN infrared SED relationship estimated using a statistically significant population size, providing future studies the ability to estimate the total infrared emission of a high-z AGN using only X-ray data. A deeper X-ray study with a similar amount multiwavelength IR data and de-absorbed X-ray luminosities would be needed to confirm this relationship is complete to lower luminosity X-ray AGN at $z>0.1$.
 
   The median infrared AGN contribution across all sources is 8\,-\,30\%, indicating that roughly 70\,-\,90\% of infrared light from this set of galaxies is coming from star formation processes. When restricting our sample to the 337 ULIRGs (L$\mathrm{_{IR}} > 10^{12}\,\mathrm{L_{\odot}}$), we find a median $f\mathrm{_{AGN}} = 13\%$, similar to the fraction found in \citet{nardini} for local ULIRGs. Looking at the median composite, rest-frame $f\mathrm{_{AGN}} $ as a function of wavelength in Figure \ref{fagn} (right), we find that AGN contribution heavily affects the mid-IR, with a maximum of nearly 80\% at 5-6\,$\mu$m. While the impact of AGN contribution trails off at wavelengths greater than $\sim$30\,$\mu$m in Figure \ref{fagn} (similar to other results, e.g. \citet{kirk12,mullaney}), the $f\mathrm{_{AGN}} $ sample distribution is broad at each wavelength with an average scatter of $\pm20-30\%$, implying that multi-component SED analysis is crucial in accurately determining the true AGN contribution for individual sources, particularly for cases without X-ray observations to constrain L$\mathrm{_{AGN}^{IR}}$. 
 
	On average, $f\mathrm{_{AGN}}$ increases with increasing X-ray and 24\,$\mu$m luminosity, but not with total infrared luminosity (see Figure \ref{xratios} and Figure \ref{ratios}). The latter tells us that any trends found with L$\mathrm{_{AGN}^{IR}}$ are not driven simply by the host galaxy\textquoteright s luminosity; or, in other words, a broad range of infrared AGN fractions can be found embedded in variously luminous galaxies. As expected, for the IR-only galaxies (represented by grey x\textquoteright s) we see fairly low AGN fractions at low and average luminosities; yet, at higher luminosities, the IR-only galaxies and XIR AGN appear similar across the log$(f\mathrm{_{AGN}}$) -- log$($L$_{24\,\mu m})$ relationship (possibly due to incomplete sample selection effects and/or SED modeling degeneracies for SMGs high redshifts, e.g. \citet{maghighz}). The 128 IR-only galaxies with IRAC colors indicative of embedded luminous AGN (not highlighted; see Section \ref{samplesel} for sample definition) span a similar range of 24\,$\mu m$ luminosities and follow nearly exactly the same trends as the XIR sample. We also considered the relationship between $f\mathrm{_{AGN}}$ and host galaxy stellar mass; again, we find a flat, non-existent correlation. We note that this sample occupies a host galaxy stellar mass distribution similar to those found in literature for AGN host galaxies, with a mean stellar mass of log(M$_{*}$)$ = 10.83\pm0.58$ \citep[e.g.][]{hickox09, xue}.
  
  We determine a clear but weak relationship in log$(f\mathrm{_{AGN}}$) -- log$($L$_{24\,\mu m})$ and in log($f\mathrm{_{AGN}}$) -- log(L$\mathrm{_{X}}$), both with slopes $\approx$0.11. Both correlations have large intrinsic scatters and weak correlation coefficients at $\sim\,\pm$60\% and $\sim$0.36, respectively. \citet{ciesla} shows that $f\mathrm{_{AGN}}$ predictions below 20\% are accompanied with large uncertainties and therefore should be disregarded; these uncertainties vary across AGN types and it is unclear how they might vary across AGN luminosities. To investigate whether there is a stronger relationship present in our more certain $f\mathrm{_{AGN}}$ calculations, we restrict our sample to $f\mathrm{_{AGN}} \geq 20\%$, which is about 28\% of the entire sample with an average log(L$_{24\,\mu m}) = 11.56 \pm0.66$ and log(L$\mathrm{_{X}}) = 44.30 \pm0.54$. Instead, we find an even weaker slope at $\sim$0.06 with a correlation coefficient of $\sim$0.12 for both sample populations, and again a large range of values. These results directly indicate a need for individual SED decomposition to infer the fraction of infrared output attributed by an AGN.

\section{Summary and Conclusions} \label{summary}
We explored the relationship between AGN activity and host galaxy dust properties across the tail end of peak AGN and galaxy growth in the Universe (redshifts $0.2 < z < 5$) using \textit{Chandra}, \textit{Herschel}, \textit{Spitzer} and NOAO Telescope observations in the Bo\"{o}tes field. We successfully disentangled AGN and star formation radiative processes in the infrared spectrum for 703 IR bright X-ray AGN using multi-component SED fitting code, \textsc{sed3fit} \citep{berta}, and determined the AGN-corrected integrated rest-frame infrared luminosity attributed to star formation, total infrared AGN luminosity, AGN dust covering factors and AGN fractions. Our main results can be summarized as follows:

\begin{itemize}

\item	We find flat trends consistent with other literature when averaging L$\mathrm{_{IR}^{SF}}$ in bins of bolometric AGN luminosity for less powerful AGN (L$_\text{AGN} < 10^{45} \,$erg\,s$^{-1}$), as well as the stronger correlations found when averaging L$_\text{X}$ in bins of star formation activity for AGN at low redshifts ($0.1 < z < 0.8 $). 
  
\item	We further decompose AGN SEDs to isolate the dusty torus component in the IR and compare to the bolometric AGN luminosity to estimate nuclear obscuration. We determine an average dust covering factor slightly lower than other literature at CF$=33\%$, which indicates a Type 2 (obscured) population of roughly a third. Further investigation of X-ray hardness reveals several X-ray AGN with covering factors that contradict the expected nuclear obscuration determined by hardness ratios (e.g. high covering factor with a low hardness ratio that is indicative of an unobscured central engine), providing further evidence that observational differences between AGN types are not primarily driven by line-of-sight dusty torus inclination.

\item	We uncover a wide range in the fraction of infrared luminosity attributed to AGN activity across all redshifts, and determine no statistically significant trend exists when evaluating $f\mathrm{_{AGN}}$ as a function of total infrared, X-ray or 24\,$\mu$m luminosity. The mean $f\mathrm{_{AGN}}$ as a function of rest-frame IR wavelength shows peak AGN contamination lives in the mid-IR range and becomes insignificant at wavelengths larger than $\sim30\,\mu$m, but the sample dispersion is large ($\pm 20-30\%$) at all wavelengths. These results demonstrate the importance of SED decomposition for individual AGN and host galaxies in order to accurately quantify AGN contamination in the IR, particularly prior to using IR photometry to estimate host galaxy properties.

\end{itemize}
While considering all of the implications discussed in this paper, we should remember that current FIR detections of intermediate and high redshift X-ray AGN in star-forming galaxies are limited by the sensitivity of far-infrared and submillimeter observatories like the \textit{Herschel} Space Observatory. The currently available resolutions limit us to the most powerful star-forming systems and we need deeper, more sensitive observations to capture the dust properties of AGN that reside in smaller and/or quiescent galaxies in order to complete the evolutionary picture. 

\acknowledgments

AB would like to thank Susan Terebey for support and advisement. This research was supported in part by NASA grants NNX15AQ06A and NNX16AF38G, research from \textit{Hubble} Space Telescope programs HST-GO-14083.002-A and HST-GO-13718.002-A, and NSF grant AST-131331. This research has made use of data from HerMES project. HerMES is a \textit{Herschel} Key Programme utilizing Guaranteed Time from the SPIRE instrument team, ESAC scientists and a mission scientist. The HerMES data was accessed through the \textit{Herschel} Database in Marseille (HeDaM - http://hedam.lam.fr) operated by CeSAM and hosted by the Laboratoire d'Astrophysique de Marseille. HerMES DR3 was made possible through support of the \textit{Herschel} Extragalactic Legacy Project, HELP (http://herschel.sussex.ac.uk).

\bibliographystyle{yahapj}
\bibliography{references}

\newpage
\section*{Appendix - Sample Extreme SEDs}
\begin{figure}[h]
\begin{tabular}{ll}
\includegraphics[width=0.5\textwidth]{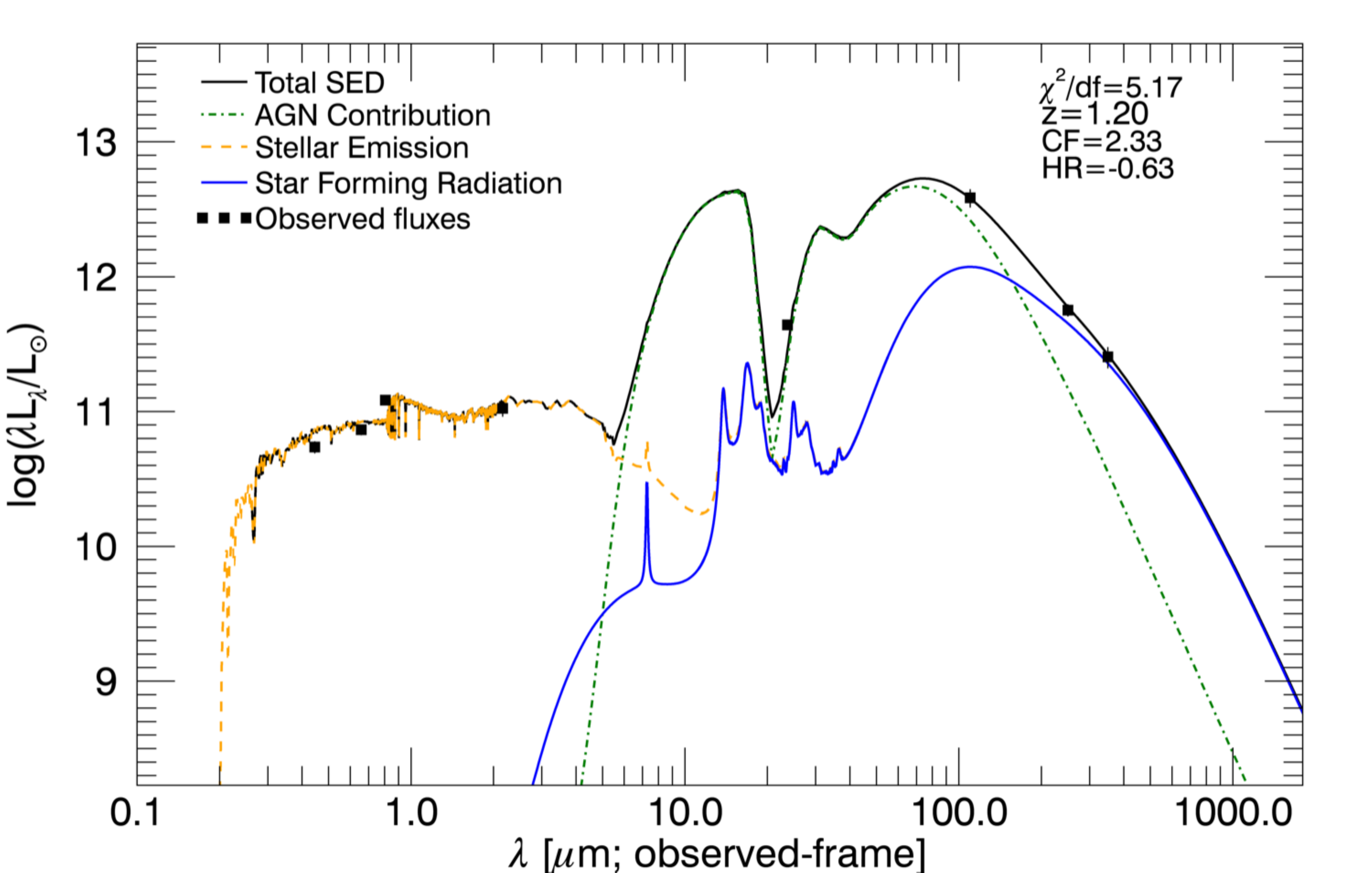}
&
\includegraphics[width=0.5\textwidth]{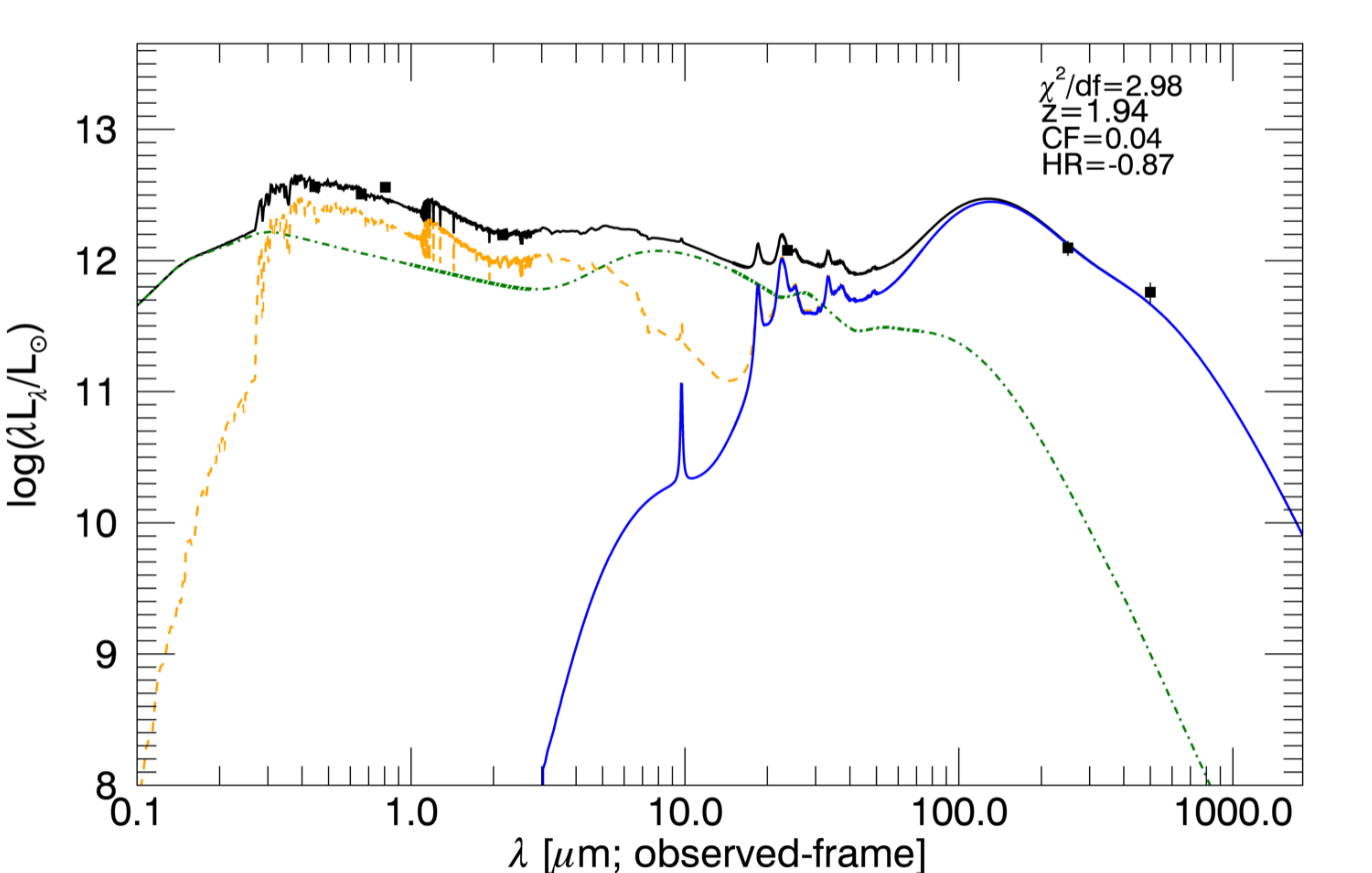}
\end{tabular}
\caption{Generated example 'extreme' spectral energy distributions of AGN with hardness ratios $\lesssim$ -0.5, indicative of an unobscured nucleus with with little to no obscuring dust and gas. \textit{Left}: SED of an X-ray unobscured AGN with high covering factor. \textit{Right}: SED of an X-ray unobscured AGN with a low covering factor.}
\label{type1}
\end{figure}

\begin{figure}[h]
\begin{tabular}{ll}
\includegraphics[width=0.5\textwidth]{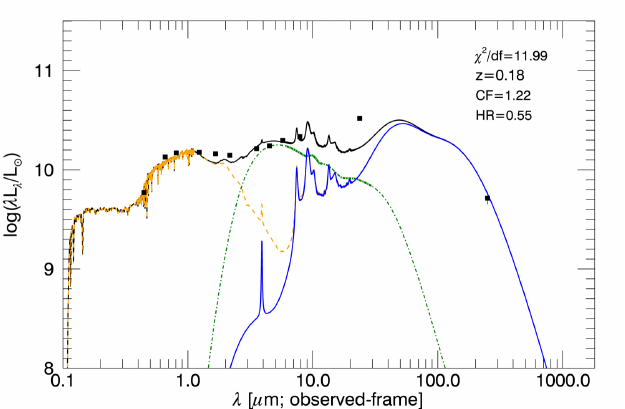}
&
\includegraphics[width=0.5\textwidth]{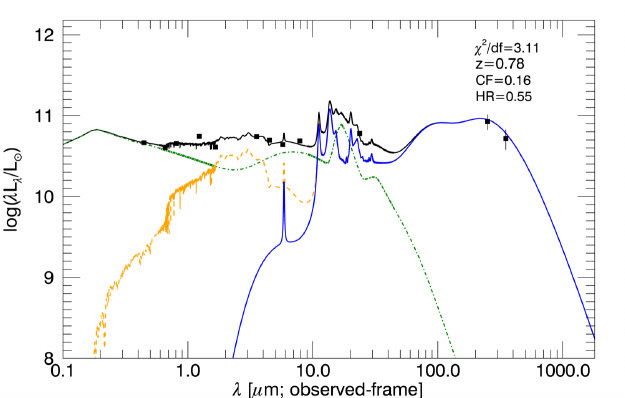}
\end{tabular}
\caption{Generated example 'extreme' spectral energy distributions of AGN with hardness ratios $\gtrsim$ 0.5, indicative of an obscured nucleus with thick absorbing circumnuclear gas and dust. \textit{Left}: SED of an X-ray obscured AGN with high covering factor. \textit{Right}: SED of an X-ray obscured AGN with a low covering factor.}
\label{type2}
\end{figure}

Below are sample SEDs of AGN that reside in the more extreme regions of Figure \ref{hrs}, marked by black crosses. Figure \ref{type1} shows two objects with X-ray HRs indicative of an unobscured nucleus with little to no dust or gas absorbing their X-ray luminosities. However, the object on the left has a high dusty torus covering factor, which is contradictory to what we'd expect to see based on the HR and the inclination-based AGN unification model \citep{antonucci,urry}. Similarly, in Figure \ref{type2} we see two objects with HRs that signify the presence of highly obscuring column densities, but SED decomposition for the object on the right determined a low covering factor that contradicts the HR estimate. These objects support the need for a different perspective on what truly drives the observational differences between AGN classifications.

\end{document}